\documentclass{APS}
\usepackage{graphicx}
\usepackage{epstopdf, epsfig}
\usepackage{siunitx}
\usepackage{booktabs}
\usepackage{multirow}
\usepackage{color}
\usepackage{array}
\usepackage{amssymb, amsmath}
 \usepackage{wasysym}
 \usepackage{rotating} 
 \usepackage{lscape}
\newcommand*\diff{\mathop{}\!\mathrm{d}}
\newcommand{\erf}{\mathrm{erf}\,}
\newcommand{\at}[2][]{#1|_{#2}}

\shorttitle{Microdroplets nucleation by dissolution of a multicomponent drop}
\shortauthor{Tan et al.}

\title{Microdroplets nucleation by dissolution of a multicomponent drop in a host liquid}

\author{Huanshu Tan\aff{1}
  \corresp{\email{huanshutan@gmail.com}},
    Christian Diddens\aff{1},
  Ali Akash Mohammed\aff{1},
     Junyi Li\aff{1},
  Michel Versluis\aff{1},
  Xuehua Zhang\aff{2,1}
  \corresp{\email{xuehua.zhang@ualberta.ca}},
  \and Detlef Lohse\aff{1,3}\corresp{\email{d.lohse@utwente.nl}}}

\affiliation{\aff{1}Physics of Fluids Group, Max-Planck-Center Twente for complex fluid dynamics, Mesa+ Institute, and  J. M. Burgers Centre for Fluid Dynamics, Department of Science and Technology, University of Twente, P.O. Box 217, 7500 AE Enschede, The Netherlands
\aff{2} Department of Chemical \& Materials Engineering, University of Alberta, Edmonton, Alberta, T6G1H9, Canada
\aff{3} Max Planck Institute for Dynamics and Self-Organization, Am Fa{\ss}berg 17, 37077 G\"ottingen, Germany}

\begin{document}

\maketitle

\begin{abstract}
Multicomponent liquid drops in a host liquid are very relevant in various technological applications.
Their dissolution or growth dynamics is complex.
Differences in solubility between the drop components combined with the solutal Marangoni effect and natural convections contribute to this complexity, which can be even further increased in combination with the ouzo effect, i.e., the spontaneous nucleation of microdroplets due to composition-dependent miscibilities in a ternary system. 
The quantitive understanding of this combined process is important for applications in industry, particularly for modern liquid-liquid microextraction processes.
In this work, as a model system, we experimentally and theoretically explore water/ethanol drops dissolving in anethole oil.
During the dissolution, we observed two types of microdroplets nucleation, namely water microdroplet nucleation in the surrounding oil at drop midheight and oil microdroplet nucleation in the aqueous drop, again at midheight.
The nucleated oil microdroplets are driven by Marangoni flows inside the aqueous drop and evolve into microdroplets rings.
A one-dimensional multiphase and multicomponent diffusion model in combination with thermodynamical equilibrium theory is proposed to predict the behavior of spontaneous emulsification, i.e. the microdroplet nucleation, that is triggered by diffusion.
A scale analysis together with experimental investigations of the fluid dynamics of the system reveals that both the solutal Marangoni flow inside the drop and the buoyancy-driven flow in the host liquid influence the diffusion-triggered emulsification process.
Our work provides a physical understanding of the microdroplet nucleation by dissolution of a multicomponent drop in a host liquid. 
\end{abstract}


\section{Introduction}

Multicomponent drops immersed in another liquid occur in a widespread range of engineering applications, such as chemical waste treatments, the separation of heavy metals, food processing, diagnostics and so on \citep{kula1982purification,fukumoto2005room,chasanis2010investigation,ahuja2000handbook,lu2017dissolution}.
In recent years, the interest in the diffusive dynamics of multiphase fluid systems has surged, as the quantitative understanding of the process is crucial not only for fundamental studies of multiphase systems, but also for its common applications in the chemical industry, particularly for modern liquid-liquid microextraction processes \citep{lohse_2016,Jain2011}.

Diffusion processes, i.e. the movement of species down a concentration gradient, can induce a mass transfer between different phases.
A classical theory about a single component bubble dissolving into a surrounding liquid was established by \citet{EpsteinPlesset1950} and later extended to drops \citep{duncan2006microdroplet,su2013mass}.
It can be derived that the radius of the bubble surface is proportional to the square root of the time, which agrees with experimental measurements.  
For multicomponent drops, the situation is very different, as the consideration of the mutual interaction of the species is necessary.
A consistent theory for the dissolution or growth of multicomponent drops in a host liquid was proposed by \citet{Chu2016}.
They employed thermodynamic equilibrium constraints at the interface with the adoption of the UNIQUAC model, which is frequently applied in the description of phase equilibria.
Molecular dynamics simulations were performed by \citet{Shantanu2017} with the conclusion of the importance of the interaction between the drop constituents and the host liquid during multicomponent drop dissolution.
All these studies focus on the investigation of pure diffusion processes.
In practice, however, the flow motion induced by the diffusion processes cannot be neglected, as the flow is able to affect the diffusion processes in turn.
A small droplet, for example, can be self-propelled by Marangoni stress when the viscosity ratio of the droplet liquid to the surrounding liquid is smaller than the length scale ratio of the droplet size to the solutal interactive length scale \citep{Ziane2014}.
\citet{Erik2016} experimentally demonstrated the existence of a transition Rayleigh number for the dissolution of a sessile multicomponent drop, above which the buoyancy-driven convection in the host liquid prevails over the diffusion.
Additionally, \citet{Dietrich2016} found that diffusion is able to induce a local concentration difference and thereby cause the segregation of the components inside the drop.

For a specific category of multiphase systems with a metastable phase regime, the diffusion phenomena are even more interesting and complex, as the phase equilibrium can be altered by the diffusion process, leading to the occurrence of metastable dispersions in the bulk \citep{solans2016spontaneous}.
Ouzo, an alcoholic beverage from Greece, is a typical example of this kind of solution.
It mainly consists of ethanol, water and (anise) oil, and it is well mixed when the oil concentration in the solution is lower than the oil solubility of the water-ethanol solvent (aqueous phase).
Spontaneous emulsification, the process of creating metastable liquid-liquid dispersions (nano- or micro-droplets), can be achieved by increasing the water concentration and thereby reducing the oil solubility without an external energy input. This is the well-known ouzo effect, which can either be triggered by simply adding water to the system or, alternatively, by the reduction of the ethanol amount by a preferential dissolution or evaporation. The latter process, i.e. the evaporation of an ouzo drop, is extremely rich and exhibits multiple phase transitions during the drying, as recently discovered by \citet{tanouzo2016,tanouzo2017}.

Due to the non-uniform evaporation rates along the liquid-gas interface and the different volatilities of water and ethanol, regions of oil supersaturation, i.e. regions where the emulsification takes place, are generated locally in the evaporating ouzo drop.
The same principle is expected to apply to a dissolving multicomponent drop, since it obeys the same dynamical equations and similar boundary conditions as the evaporating ouzo drop.

\begin{figure}
\centering{
\includegraphics[width=1\textwidth]{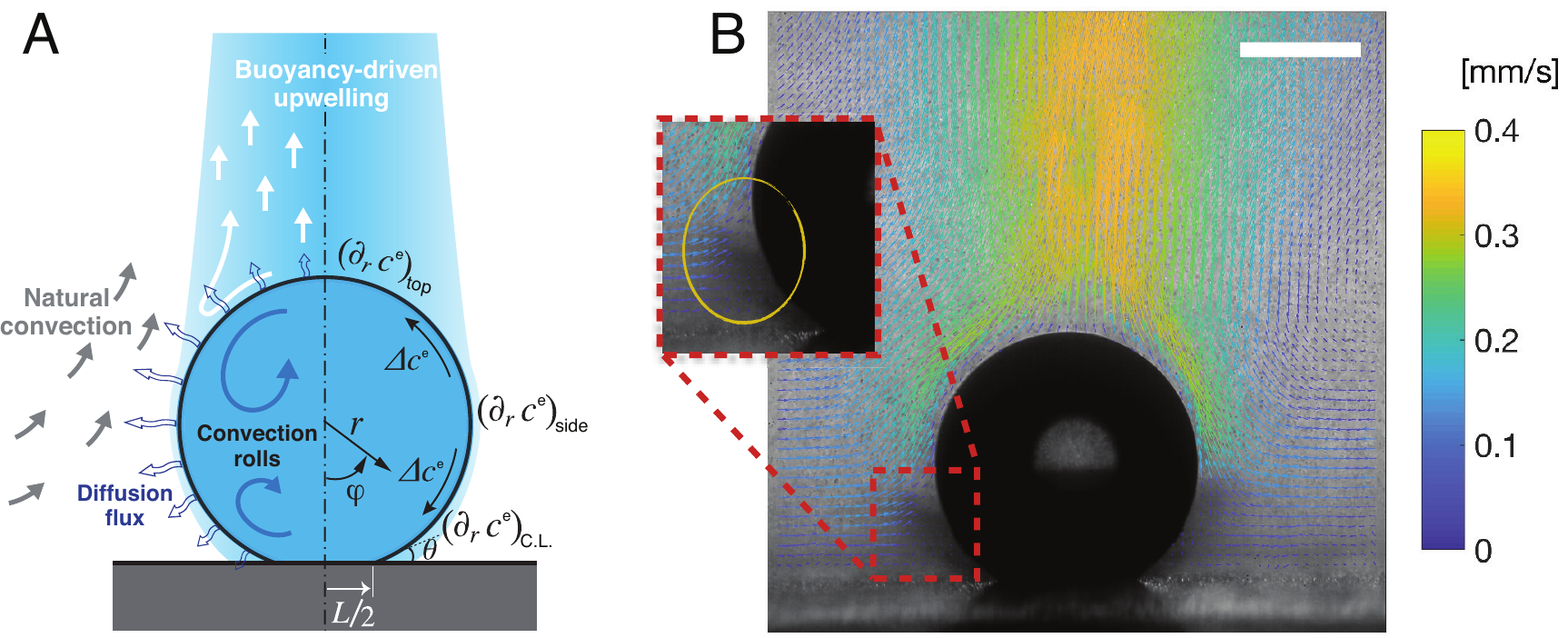}}
\caption{(A) Hydrodynamic sketch of a dissolving sessile drop.
The left side presents the flow directions and diffusion fluxes along the drop surface. 
Geometrical and physical quantities are defined on the right side.
(B) Experimental snapshot of the drop overlaid with the external flow field obtained by PIV-measurement.
The scale bar is \SI{0.45}{\milli\meter}.}
\label{fig:sk}
\end{figure}

In this paper, we explore the dissolution of a multicomponent drop in a host liquid (Fig. \ref{fig:sk}), with a particular focus on systems which have the capacity of undergoing the spontaneous emulsification (ouzo effect).
Experimental steps and methods are discussed in Section \ref{sec:Emethod}.
The sessile drop picked here consists of the two miscible components water and ethanol with different initial ratios, and anethole oil acts as host liquid, in which ethanol is miscible and water is immiscible.
The general observations and descriptions about the dissolution phenomena are given in Section \ref{sec:Dissolution process}.
During the dissolution, we observed both spontaneous emulsifications in the region of the host oil (oil-rich phase) and in the sessile drop region (aquous phase), i.e. water-in-oil (w-in-o) microdroplets and oil-in-water (o-in-w) microdroplets correspondingly, which are presented in Section \ref{sec:emu}.
In Section \ref{sec:oneDModel}, we develop a one-dimensional diffusion model with the adoption of the so-called diffusion path theory \citep{kenneth1972} and thermodynamical equilibrium theory to provide insight to the occurrence of the spontaneous emulsification and its evolution.
The so-called UNIFAC model was applied to describe phase equilibria with the consideration of liquid activity coefficients.
An alternative model would be the so called UNIQUAC model used by \citet{Chu2016}, but the parameters of the mixture in this study are not available for that model. 
Through this model, we gain insight into the emulsification process (diffusion-induced microdroplet nucleation), as well as the mass transport caused by the pure diffusion process, which are presented and evaluated in Section \ref{sec:prediction}.
To figure out the influence of flow motions on the emulsification process,
a scaling analysis and micro-PIV measurements were performed in Section \ref{sec:hydrodynamics}.
The scaling analysis reveals that the Marangoni effect dominates the flow motion inside the drop, while natural convection is dominant in the host liquid (Fig. \ref{fig:sk}A), which was confirmed by side recording movies and micro-PIV measurement results (Fig. \ref{fig:sk}B).
Having obtained a good understanding of the fluid dynamics in the system, we finally acquire a more systematic understanding about the dissolution process and the preferred position where the diffusion-triggered emulsification takes place.

\section{Experimental method}
\label{sec:Emethod}
\subsection{Solution and substrate}
The system we investigate here consists of Milli-Q water [obtained from a Reference A$+$ system (Merck Millipore) at \SI{18.2}{\mega\ohm} at \SI{25}{\degreeCelsius}], ethanol (Sigma-Aldrich; $\geqslant\SI{98}{\percent}$), and anethole oil (Sigma-Aldrich; trans-anethole, $\geqslant\SI{99.8}{\percent}$).
We performed dissolution experiments in a cuvette (Hellma; Inner dimensions $\SI{30}{\milli\meter}\times \SI{30}{\milli\meter}\times \SI{30}{\milli\meter}$), on the bottom of which a hydrophobized glass slide ($\approx \SI{20}{\milli\meter}\times \SI{20}{\milli\meter}$) was placed.
A certain amount of water-saturated anethole was added into the cuvette, performing the host liquid.
The depth of the liquid was \SI{7.5}{\milli\meter}.
Water-ethanol binary drops with different volumetric concentrations of ethanol (\SI{30}{\percent}, \SI{40}{vol\percent}, \SI{50}{vol\percent}, \SI{60}{vol\percent}, \SI{70}{vol\percent}) were produced in the oil through a custom needle (Hamilton; outer diameter/inner diameter \SI{0.21}{\milli\meter}/\SI{0.11}{\milli\meter}) by a motorized syringe pump (Harvard; PHD 2000), and then directly deposited on the centre of the hydrophobized glass surface.

\subsection{Emulsion/microdroplets recognition}
The emulsions (nucleated microdroplets) were recognised visually.
Both the water-in-oil (w-in-o) emulsion and the oil-in-water (o-in-w) emulsion had a recognisable cloudy-white appearance because of the microscopic size of the nucleated microdroplets, which enable them to scatter all the colours equally. 
The recognition of the emulsions was processed by watching the recorded videos frame by frame. 
Although dissolving drops were in millimeter-scale, their spherical shapes with high curvature unavoidably caused reflected light spots, which increased the difficulty of the recognition of the presence of the microdroplets inside the drop.
Therefore, the presence or absence of emulsions were carefully determined by detecting the liquid colour variation and their movement from the recorded videos (both top views and synchronised  side views).
No fluorescence technique were used for the microdroplet detection, in order to avoid any influence of the added fluorescent materials on the spontaneous emulsification.

\subsection{Micro-PIV}
To investigate the flow field around the dissolving drop, we added tracking particles (Dantec dynamics; PSP-5, diam. \SI{5}{\micro\meter}, made by nylon-12) in the host liquid at a seeding density of \SI{0.2}{\milli\gram\per\milli\liter} to perform micro-PIV measurements.
As shown in appendix \ref{sec:A1} these particles can be considered as passive.
Thanks to the low flow rate in our study, a continuous LED light source (Thorlabs; MCWHL5) was able to provide enough volume illumination for the measurements.
The light source and the camera were placed at two opposite sides of the cuvette.
The light passed through convex lenses before illuminating the cuvette to form a parallel light beam to increase the image contrast.
At the other side, we positioned a high-speed camera [Photron Fastcam SA2 32GB, 50 frames per second(fps) at $2,048 \times 2,048$ pixel resolution] attached with the microscope system (Infinity; Model K2 DistaMax) to perform high-speed imaging.
The position of the recording system was adjusted to have a focal plane crossing the droplet centre.
The thickness of the focal plane is \SI{0.02}{\milli\meter}.
Thereby, we could obtain sharp images of the tracking particles within a cross-sectional plane of the drop.
We took image pairs with an inter-framing time of \SI{20}{\milli\second} every two seconds.
The obtained image pairs were first processed to reduce the noise, and then imported into PIVlab software \citep{thielicke2014pivlab, williamPHD} to calculate the flow field. 
The size of interrogation window was taken as a $128\times128$ pixel matrix, corresponding to $\SI{142}{\micro\meter}\times\SI{142}{\micro\meter}$. 
The interrogation window overlap was set as \SI{75}{\percent}, leading to a \SI{35.5}{\micro\meter} vector spacing in the calculated velocity matrix.

\section{Dissolution process}
\label{sec:Dissolution process}
\subsection{Characteristic states of dissolving drops}
\begin{figure}
\centering{
\includegraphics[width=0.76\textwidth]{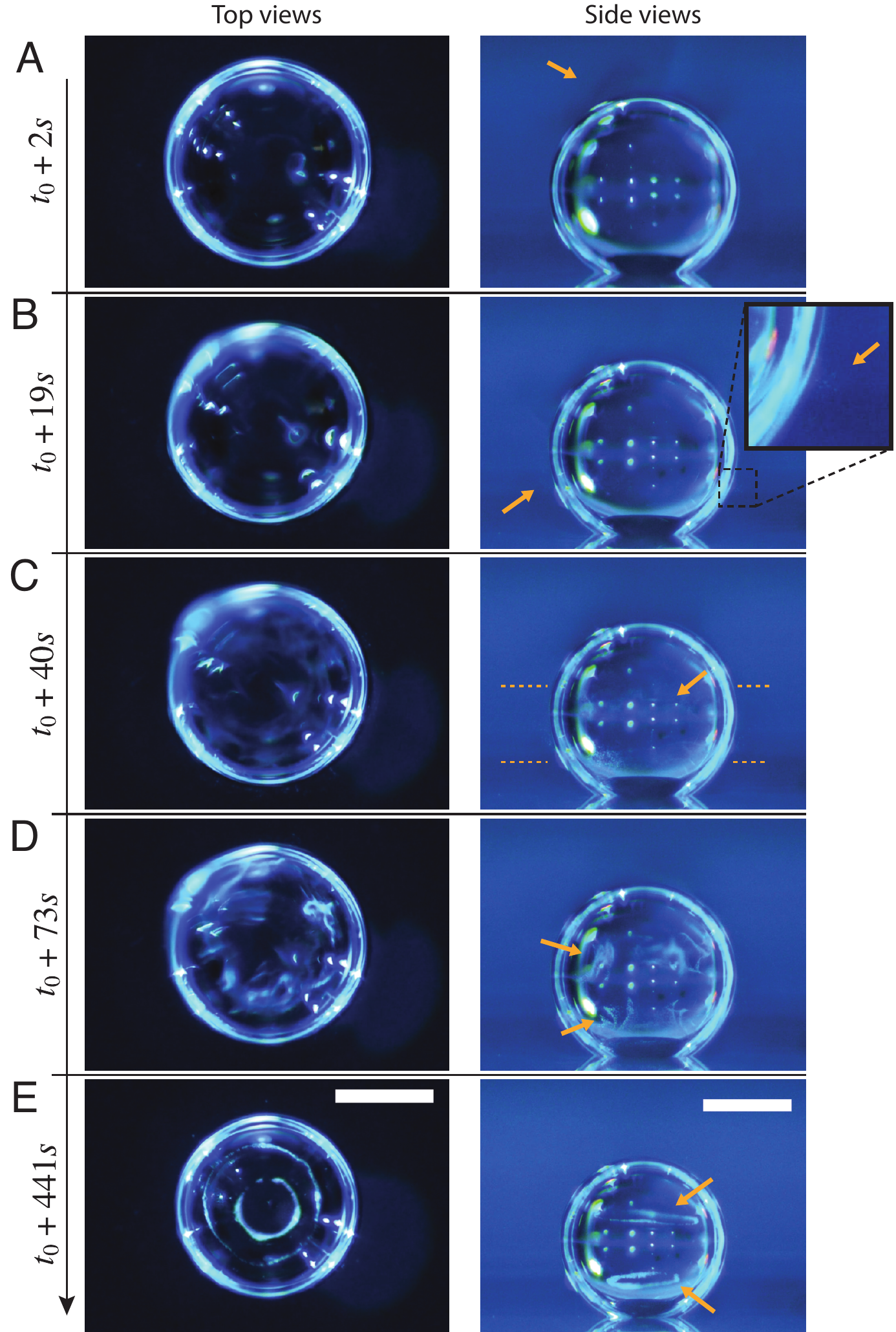}}
\caption{Experimental snapshots of a dissolving water/ethanol (v/v 40:60) drop in anethole oil experiencing different characteristic states during the dissolution. 
The first column of photos are top views and the second column are the corresponding side views.
The initial drop size is around $\SI{0.5}{\micro\liter}$ and $t_0$ denotes the drop deposition time.
(A) The drop begins with a transparent appearance surrounded by a clean host liquid.
The arrow indicates solute plume above the drop.
(B) W-in-o emulsion (water microdroplets) suspends outside the drop. The arrows and the inserted zoom-in panel highlights the location of the microdroplets.
(C) O-in-w emulsion (oil microdroplets) appears inside the drop with a preferential location around the equator of the drop (arrows, area in between the two horizontal lines).
(D) More oil microdroplets are formed and concentrate at the two sides of the equator. The w-in-o emulsion disappears.
(E) In around 7 minutes, the two concentrated clusters of microdroplets evolve into two rings. 
The scale bar is \SI{0.5}{\milli\meter} in all those figures. \textcolor{blue}{(Corresponding movie: supplementary material Movie S2 and Movie S3)}
}
\label{fig:snap}
\end{figure}

A dissolving water-ethanol (aqueous) sessile drop in anethole oil (host liquid) is displayed in Figure \ref{fig:snap}. The initial ethanol concentration is \SI{60}{vol\percent} and the initial drop volume is around \SI{0.5}{\micro\liter}.
The experimental snapshots (the left column are the top views and the right column are the corresponding side views) present several interesting phenomena occurring during the drop dissolution process, including an upwards rising solute plume, detaching from the top of the drop (arrows in Fig. \ref{fig:snap}A), spontaneous emulsifications inside and outside the drop (arrows in Fig. \ref{fig:snap}BCD), and double oil-microdroplets rings forming by convection rolls and suspended in the drop (arrows in Fig. \ref{fig:snap}E).

At the beginning when the drop was deposited on the hydrophobic substrate in the oil host liquid, both the drop and the surrounding oil are transparent, except for the shadows above the drop, as displayed in Figure \ref{fig:snap}A.
The shadow, indicated by the arrow in the figure, is the solute plume, i.e. the ethanol-rich oil mixture as ethanol diffuses into the oil surrounding the drop.

In our system, there are two kinds of self-emulsifications: water-in-oil emulsification and oil-in-water emulsification.
The former one creates water microdroplets in the oil host liquid (w-in-o emulsion) and the latter one generates oil microdroplets in the aqueous drop (o-in-w emulsion).
The first appearing microdroplets are the w-in-o emulsion ones. They nucleate at a certain location in the surrounding oil, comparable to the position of the earth's tropic of capricorn.
The emulsification normally starts within less than \SI{5}{\second} after drop deposition.
Figure \ref{fig:snap}B is a snapshot taken at \SI{19}{\second} to provide a visualisation of the emulsion and the inserted zoom-in panel highlights their position, indicated by arrows. 
The microdroplets suspend in oil for a while and then disappear. 
Around half a minute later, o-in-w emulsification sets in inside the drop, preferentially in the middle of the drop, concentrating in the region pointed at by the arrow in Figure \ref{fig:snap}C.
Notable, the preferential location is different from that reported in our previous work on evaporating drops, in which droplet nucleation either occurred at the contact line region for flat evaporating ouzo drops \citep{tanouzo2016}, or at the top of the drop for spherical evaporating ouzo drops \citep{tanouzo2017}.
More detailed discussion about those spontaneous emulsifications will be given in Section \ref{sec:emu}.

Yet another remarkable phenomenon is that the generated o-in-w emulsions gradually form two rings of microdroplets in the drop.
The generated microdroplets firstly split into two groups, one above and one below the equator of the drop, which is shown in Figure \ref{fig:snap}D.
The mechanism of the migration comes from two Marangoni convection rolls located separately above and below the equatorial plane (\textcolor{blue}{see the supplementary material Movie S1}).
The Marangoni flow motion is induced by surface tension gradients due to concentration variations along the interface, i.e. solutal Marangoni flow.
The convection rolls drive the nucleated oil microdroplets and lead to an accumulation of these in the centre of each vortex-roll, resulting in the formation of two rings of oil microdroplets as shown in Figure \ref{fig:snap}E.
More detailed discussion on the dynamics will be given in Section \ref{sec:hydrodynamics}.
When the ethanol in the drop finally has dissolved, the solute Marangoni effect stops and the rings disintegrate (\textcolor{blue}{see the supplementary material Movie S1}).

All these interesting phenomena happen during the first stage, which is mainly driven by the dissolution of ethanol from the drop. This stage takes around \SI{30}{\minute}.
In the second stage, the remaining water diffuses with an extremely slow speed. During that stage, no further unexpected phenomena occur.
Therefore, in this paper we only investigate the first stage of the process, up to the time the alcohol has fully dissolved.
It is important to point out that, during the whole dissolution process, there is always a distinguishable interface between the oil host medium and the drop medium for all experiments (drops with different water-ethanol ratios) we performed.
The sharp boundary of the drop corresponds to near-discontinuities in the gradient of the concentration-distance curve \citep{hartley1946},
which reflects that the drop solution and host solution are macroscopically phase-separated at the drop-oil interface.
We also stress that our experiments are very reproducible, even quantitatively.

\subsection{Dissolution of drops with different initial ethanol concentrations}
\begin{figure}
\centering{
\includegraphics[width=\textwidth]{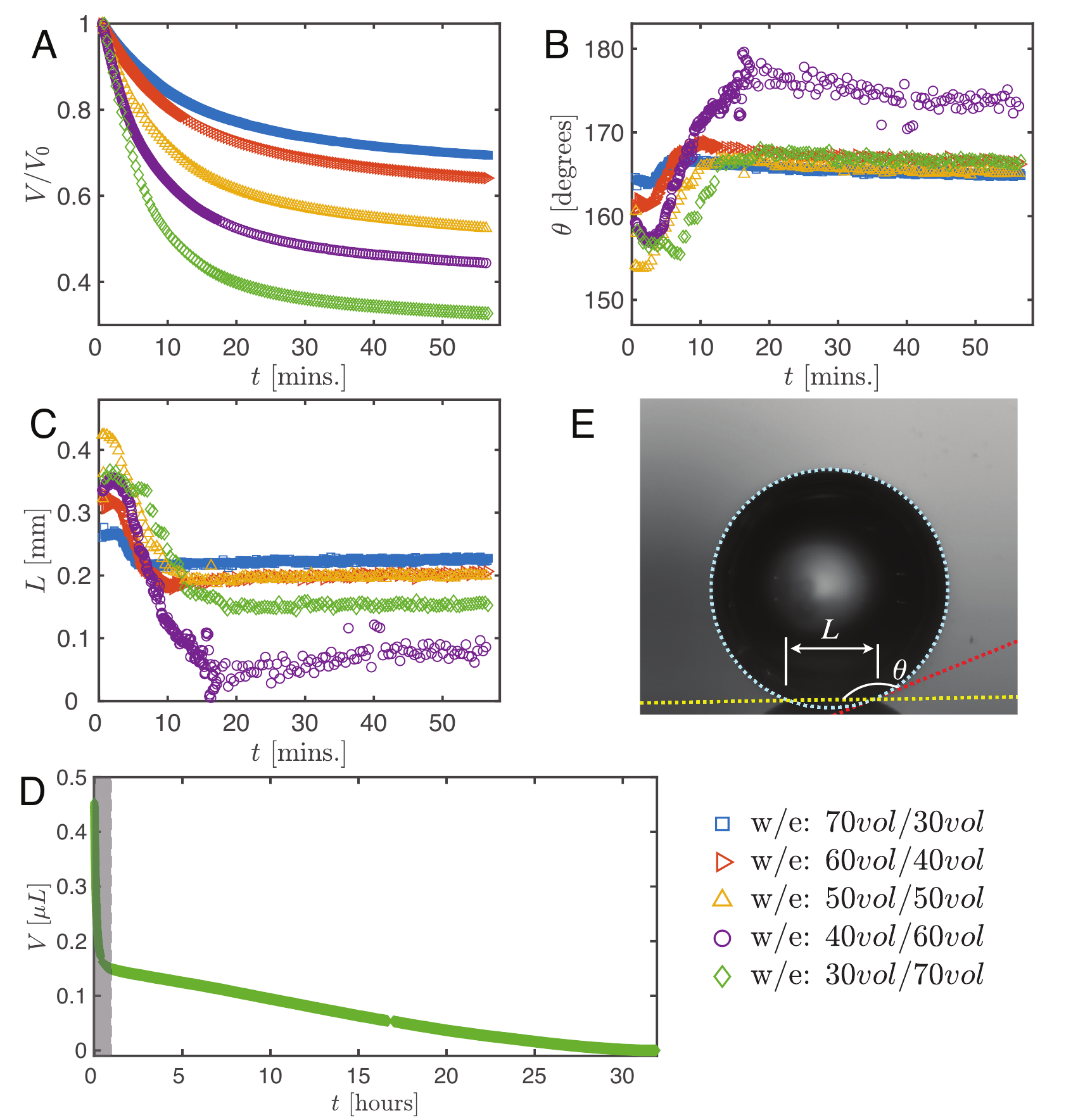}}
\caption{Morphology evolution of dissolving drops created with five different initial water-to-ethanol ratios (v/v 30:70, 40:60, 50:50, 60:40, 70:30).
(A-C) present temporal evolutions of the nondimensionalized volume $V/V_0$ by initial drop size $V_0$, contact angle $\theta$, and footprint diameter $L$, respectively, during the dissolution.
For one case (v/v 30:70) the volume evolution of the whole dissolution process is displayed in (D) long-time behavior of droplet volume.
The shaded area on the left is shown in (A). In (E), a recording image of a dissolving drop is shown with annotations of the geometrical parameters.
}
\label{fig:geo}
\end{figure}

To quantitively investigate the phenomena, we repeated the dissolution experiment with the drops in different initial water-ethanol ratios, from \SI{30}{vol\percent} to \SI{70}{vol\percent} ethanol. 
Figure \ref{fig:geo} shows the dissolution characteristics of the drops, including the temporal evolution of the drop volume, the variations of the contact angle $\theta$ and the footprint diameter $L$.
The annotations of the geometrical variables are available in a raw picture from the experiment (Fig. \ref{fig:geo}E).

The volume evolutions of the drops are nondimensionalized by the initial drop size $V_0$ to demonstrate a declining trend of the residual water volume with an increasing initial ethanol concentration, as apparent from Figure \ref{fig:geo}A.
The figure also reveals that indeed \textit{all} drops experience two stages with two distinguished dissolving rates, as discussed above: These two stages correspond successively to the initial stage dominated by the dissolution of the ethanol and the subsequent slow dissolution of the remaining water.
This is supported by the consistence between the initial water ratios and the drop residual volume percentage after the stage transition.
The same behaviour also exists in the evaporation process of multicomponent drops \citep{tanouzo2016,liu2008}.
Water has extremely small solubility in oil (immiscible), therefore, the second stage takes much longer time than the first one (Fig. \ref{fig:geo}D).

The evolutions of the contact angle (Fig. \ref{fig:geo}B) and the footprint diameter (Fig. \ref{fig:geo}C) also reveal the variation of the ethanol content in the drop.
In the first \SI{10}{} to \SI{15}{\minute}, the contact angle increases by around \SI{20}{} degrees, accompanied by a receding of the contact line.
The increase of the contact angle is a result of the rising water concentration in the drop as ethanol is dissolving much faster than water.
In this period, neither the constant contact radius (CR)-mode nor the constant angle (CA)-mode applies \citep{lohse2015rmp}.
In the second stage, the drop evolves nearly in CR-mode -- there is a slightly decreasing contact angle with a \textit{stabilized} contact area. 
During the investigated process, the drop has a very high contact angle, more than \SI{150}{\degree}, because of the higher interfacial energy between the substrate and anethole compared to the energy between the substrate and the aqueous solution.

It is worth noting that, at the very beginning ($\sim \SI{2}{\minute}$), the contact angle decrease with time and the footprint diameter increases by a small amount.
To our best knowledge, this is the first observation of this phenomenon in dissolving multicomponent drops, although there are a few reports on this phenomenon in evaporating multicomponent drops \citep{liu2008,sefiane2008}.
A plausible explanation is that after deposition of the drop, the ethanol molecules in the drop tend to move towards the surface because of the lower interfacial energy between the hydrophobic substrate and ethanol compared to water, which results in the decreasing contact angle and the increasing contact area size \citep{liu2008}.

\section{Spontaneous emulsification}
\label{sec:emu}
As stated above, during the dissolution, spontaneous emulsification happens both in the host oil and in the drop, forming w-in-o emulsions and o-in-w emulsions successively.
The former ones are nucleated water microdroplets suspending in anethole oil, whereas the latter ones are anethole oil microdroplets nucleated in the aqueous phase.
We performed 24 groups of dissolution experiments with five different water-to-ethanol ratios as the drop initial composition to study the impact of the composition on the emulsification phenomena.
In Table \ref{tab:emu}, the observed emulsification behaviours for the different initial drop compositions are listed.
Y/N stands for presence/absence of the emulsification.
The waiting time for the onset of emulsification, measured with respect to the moment of needle detachment from the drop, are given in parentheses.
W-in-o emulsification occurs only when initial ethanol content in drop is high ($\geq \SI{50}{vol\percent}$) and its onset time is very short, within seconds.
On the contrary, o-in-w emulsification occurs for \textit{all} the cases, independent of the ethanol content, but more than half a minute later.
The corresponding onset time has a \textit{negative} correlation with the initial ethanol content in the drop.
Experimental videos are available \textcolor{blue}{in the supplementary material Movie S4}.

\begin{figure}
\centering{
\includegraphics[width=0.9\textwidth]{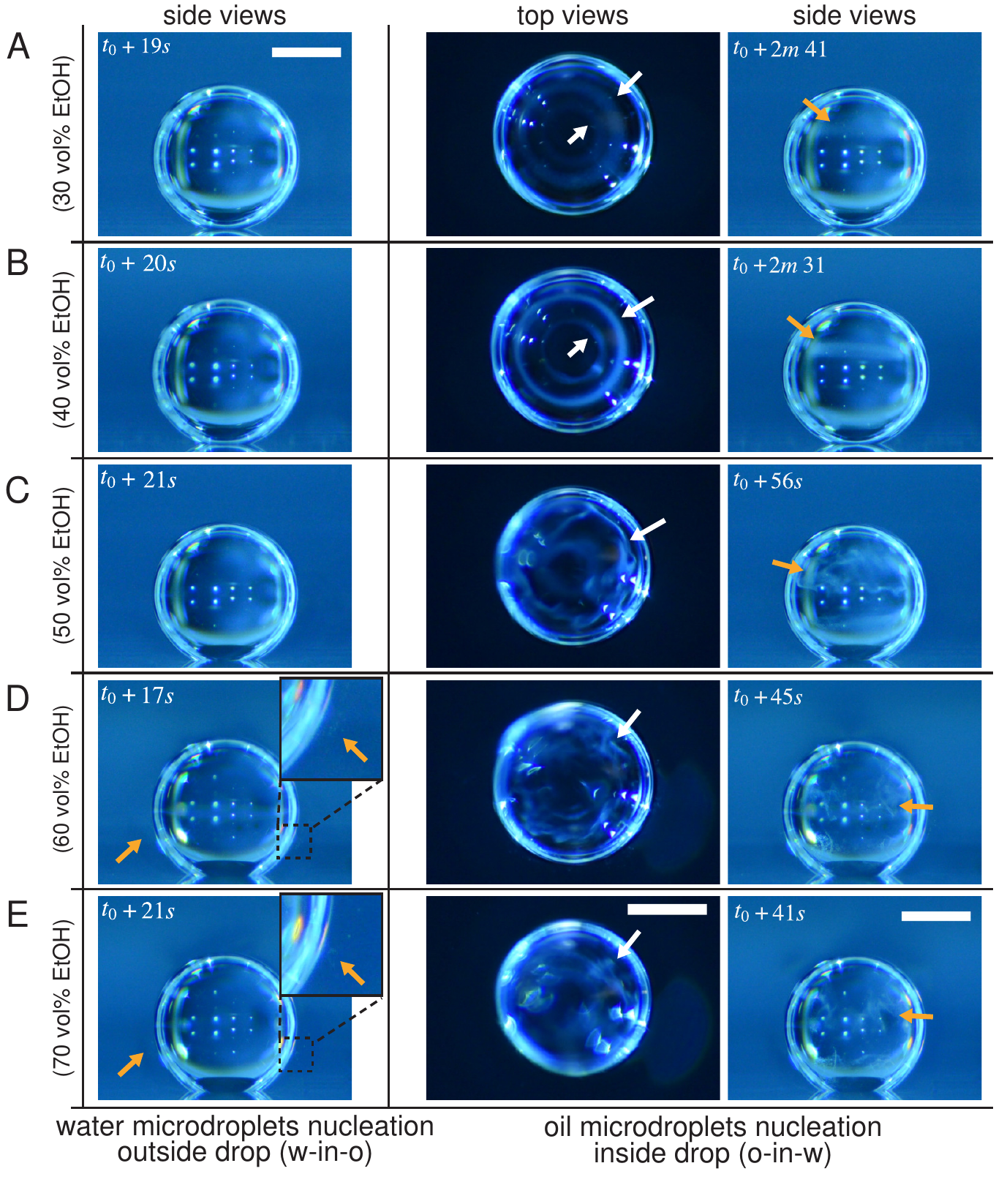}}
\caption{Experimental snapshots showing appearance or absence of emulsifications during the dissolution of water/ethanol drops with different initial water-to-ethanol volume ratios (A-E for 30:70, 40:60, 50:50, 60:40, and 70:30, respectively).
The first column of the photographs show that water-in-oil emulsification only happens for the high initial ethanol concentration cases (\SI{60}{vol\percent} and \SI{70}{vol\percent}).
The water microdroplets around drops are pointed at by black arrows in the zoom-in figures of D and E.
The last two columns of the synchronized side- and top-view photographs display the oil-in-water emulsification occurs inside drops in all the cases.
The o-in-w emulsion are pointed at by the arrows.  
The scale bars are \SI{0.5}{\milli\meter} in all these figures.
Some experimental videos are available \textcolor{blue}{in the supplementary material Movie S4}.
}
\label{fig:emu}
\end{figure}

Figure \ref{fig:emu} shows photographs of the emulsion detection experiments.
Cloudy-white w-in-o emulsions (nucleated water microdroplets) appear and suspend outside the drop, when the drop is created with \SI{60}{\percent} and \SI{70}{\percent} volume percentage ethanol (the first column of Figs. \ref{fig:emu} D and E).
While for the \SI{30}{vol\percent} and \SI{40}{vol\percent} cases, the surrounding anethole oil remains clean -- there is no w-in-o emulsions (the first column of Figs. \ref{fig:emu} AB).
\SI{50}{vol\percent} is apparently the transition point, as for this case, in some experiments the w-in-o emulsions appear and in others they do not (the first column of Figs. \ref{fig:emu} C). 
As stated above, in all of our experiments, we found that the emulsions only appear at a certain location, close to the tropic of capricorn of the drop and in the vicinity of the drop interface, as pointed out by arrows in the inserted pictures.
The appeared emulsions move up and down at this location and some microdroplets are driven away by natural convection in the host liquid (cf. section \ref{sec:hydrodynamics} and see the supplementary material Movie S2).
After around half a minute, the w-in-o emulsions disappear and the surrounding liquid becomes transparent again.

The o-in-w emulsions (nucleated oil microdroplets) in the drop show up independently of the initial ethanol concentration of the drop.
The cloudy-white o-in-w emulsions inside the drop are visible in all cases, as displayed in the second and third columns of Figures \ref{fig:emu}A-E.
The oil microdroplets emerge near the equator of the drop and then follow the Marangoni flow.
Gradually, the oil microdroplets concentrate at the centre of the convection rolls and form two microdroplets rings.
We provide top view photographs (the second column of Figs. \ref{fig:emu}A-E) for each case, from which the rings are clearly visible.
The first two rows of the photographs (the second column of Figs. \ref{fig:emu}AB) show two already formed rings of oil microdroplets (see arrows), whereas the last three rows (the second column of Figs. \ref{fig:emu}CDE) show the early chaotic arrangement of the microdroplets before the rings formed (see arrows).

\newcolumntype{P}[1]{>{\centering\arraybackslash}p{#1}}

\begin{table}
  \begin{center}
\def~{\hphantom{0}}
  \begin{tabular}{P{1.5cm}| P{.7cm} P{0cm} |P{1.5cm}  P{1.5cm}  |P{1.5cm} P{1.5cm}}
  \toprule
    \multicolumn{1}{c}{drop} & \multicolumn{1}{c}{} & \multicolumn{1}{c}{} & \multicolumn{4}{c}{spontaneous emulsification}\\[3pt]
      \cmidrule(r){1-1} \cmidrule(r){4-7 }
 \multicolumn{1}{c}{solution}  & \multirow{2}{*}{trial} &  \multicolumn{1}{c}{} & \multirow{2}{*}{w-in-o} &  \multicolumn{1}{c}{onset} & \multirow{2}{*}{o-in-w} & onset  \\
\multicolumn{1}{c}{w:e}&   \multicolumn{1}{c}{}&  \multicolumn{1}{c}{}& &  \multicolumn{1}{c}{time} & &  \multicolumn{1}{c}{time} \\
  \toprule
    \toprule
 &1 && N & - &Y & \SI{69.0}{\second}\\
7:3 &2 && N & - &Y & \SI{81.5}{\second}\\
(v/v) &3 && N & - &Y & \SI{62.0}{\second}\\
 &4 && N & - &Y & \SI{69.5}{\second}\\
 \cline{1-7}
 &  &  &  &  &  &   \\[-0.7em]
 &1 && N & - &Y & \SI{57.5}{\second}\\
6:4 &2 && N & - &Y & \SI{70.0}{\second}\\
(v/v) &3 && N & - &Y & \SI{61.0}{\second}\\
 &4 && N & - &Y & \SI{62.5}{\second}\\    
  \cline{1-7}    
   &  &  &  &  &  &   \\[-0.7em]     
 &1 && Y & \SI{5.0}{\second} &Y &\SI{27.0}{\second}\\
 &2 && N & - &Y & \SI{33.0}{\second}\\
5:5 &3 && Y & \SI{5.0}{\second} &Y &\SI{32.5}{\second}\\
(v/v) &4 && Y & \SI{3.0}{\second} &Y &\SI{36.0}{\second}\\
 &5 && Y & \SI{2.0}{\second} &Y &\SI{28.0}{\second}\\    
  \cline{1-7}    
   &  &  &  &  &  &   \\[-0.7em]     
 &1 && Y & \SI{5.0}{\second} &Y &\SI{26.0}{\second}\\
4:6 &2 && Y & \SI{3.5}{\second} &Y &\SI{29.0}{\second}\\
(v/v) &3 && Y & \SI{1.5}{\second} &Y &\SI{29.5}{\second}\\
 &4 && Y & \SI{3.0}{\second} &Y &\SI{39.5}{\second}\\
  \cline{1-7}    
   &  &  &  &  &  &   \\[-0.7em]     
 &1 && Y & \SI{4.0}{\second} &Y &\SI{28.0}{\second}\\
 &2 && Y & \SI{3.0}{\second} &Y &\SI{30.0}{\second}\\
3:7 &3 && Y & \SI{3.5}{\second} &Y &\SI{30.0}{\second}\\
(v/v) &4 && Y & \SI{5.0}{\second} &Y &\SI{29.0}{\second}\\  
 &5 && Y & \SI{2.0}{\second} &Y &\SI{25.5}{\second}\\      
 &6 && Y & \SI{3.0}{\second} &Y &\SI{29.5}{\second}\\   
 &7 && Y & \SI{1.0}{\second} &Y &\SI{34.5}{\second}\\
       \bottomrule
  \end{tabular}
  \caption{Spontaneous emulsifications by dissolution of multicomponent drops with different initial compositions in the host liquid.
The subindices e, w, and o denote components ethanol, water, and anethole oil, respectively. Y and N stands for presence and absence of the emulsification, respectively. The time values are onset time of emulsification.}
  \label{tab:emu}
  \end{center}
\end{table}

\section{One-dimensional multicomponent diffusion model}
\label{sec:oneDModel}
\subsection{Idea of the model}
More quantitative insight into the spontaneous emulsification is gained by theoretically analysing the multi-diffusion process of the water-ethanol drop dissolving in the host liquid.
A pure diffusion model is developed for \textit{the early stage} of the diffusion process, which, together with the so-called diffusion path method proposed by \citet{kirkaldy1963} and \citet{kenneth1972}, is used to predict the appearance  or absence of emulsification.
Our model consists of two parts, namely one part calculating the liquid-liquid equilibrium at the interface of the two regions (Sec. \ref{sec:LLE}) and the other part modelling the mass transport in the two multicomponent fluids in contact
 (Sec. \ref{sec:mass}).
The mass transport is modelled as a one-dimensional problem with a moving interface separating the aqueous phase and the oil-rich phase, known as Stefan problem \citep{crank1979mathematics}.
The liquid-liquid equilibrium of the ternary mixture is calculated by applying the condition of equal chemical potentials (fundamental thermodynamic relation) in combination with the UNIFAC model to quantify the nonideality of the mixture.

Mass transport in liquids happens on a large time scale compared to the macroscopic phase separation at the interface, and therefore these two processes are decoupled in our model.
To decouple them, the following assumptions were made:
First, we assume that the equilibrium at the interface is instantaneously achieved and remains stable during the mass transport across the interface.
To further simplify the model, we also assume zero boundary thickness, i.e. disregarding the microscopic details of the thermodynamic equilibrium process.
With these assumptions, the interfacial composition is directly given by the liquid-liquid equilibrium calculation and thereby determines the mass transport model.

Stefan problems commonly exists in many studies involving diffusion, such as heat transfer with a phase transition (thawing, freezing, melting), moisture transport of swelling grains or polymers \citep{barry2008}. Classic solutions to Stefan problems are given by \citet{crank1979mathematics}.
In subsection \ref{sec:mass}, we present definitions and the derivation of the equations with respect to our problem.

But before, in subsection \ref{sec:LLE}, we give a brief description of the applied equilibrium theory.
The establishment of the liquid-liquid equilibria between two phases happens on a short time scale compared to the diffusion process in the adjacent bulk regions.
Once the aqueous medium and the host liquid are in contact, thermodynamic equilibrium favours an oil-rich phase on one side, coexisting with a water-rich phase on the other side, i.e. a macroscopic phase separation, which is observed as the sharp boundary of the water-ethanol drop in oil.
The composition of these two phases are situated on the binodal curve (or coexistence curve) in the ternary phase diagram of the system and can be connected by a tie line, see Figure \ref{fig:comp}B.


\subsection{Liquid-liquid equilibrium at interface}
\label{sec:LLE}
The liquid-liquid equilibrium is achieved when the chemical potentials $\mu_\alpha$ are the same in both phases for each species $\alpha$.
The chemical potential is a function of pressure, temperature, and mole fractions $x_\alpha$ of the liquid constituents.
At fixed temperature and pressure, a three component liquid-liquid equilibrium system only has a single degree of freedom \citep{Chu2016}.
The condition of equality of the chemical potentials then reduces to
\begin{equation}
x_\alpha^{\text{d}} \, \gamma_\alpha^{\text{d}}= x_\alpha^{\text{h}} \, \gamma_\alpha^{\text{h}},
\label{eqn:cp}
\end{equation}
where $\gamma$ is the liquid activity coefficient, a correction factor accounting for the nonideality of the mixture.
The superscripts indicate the region, i.e. either drop region (d) or host liquid region (h). 

\begin{figure}
\centering{
\includegraphics[width=\textwidth]{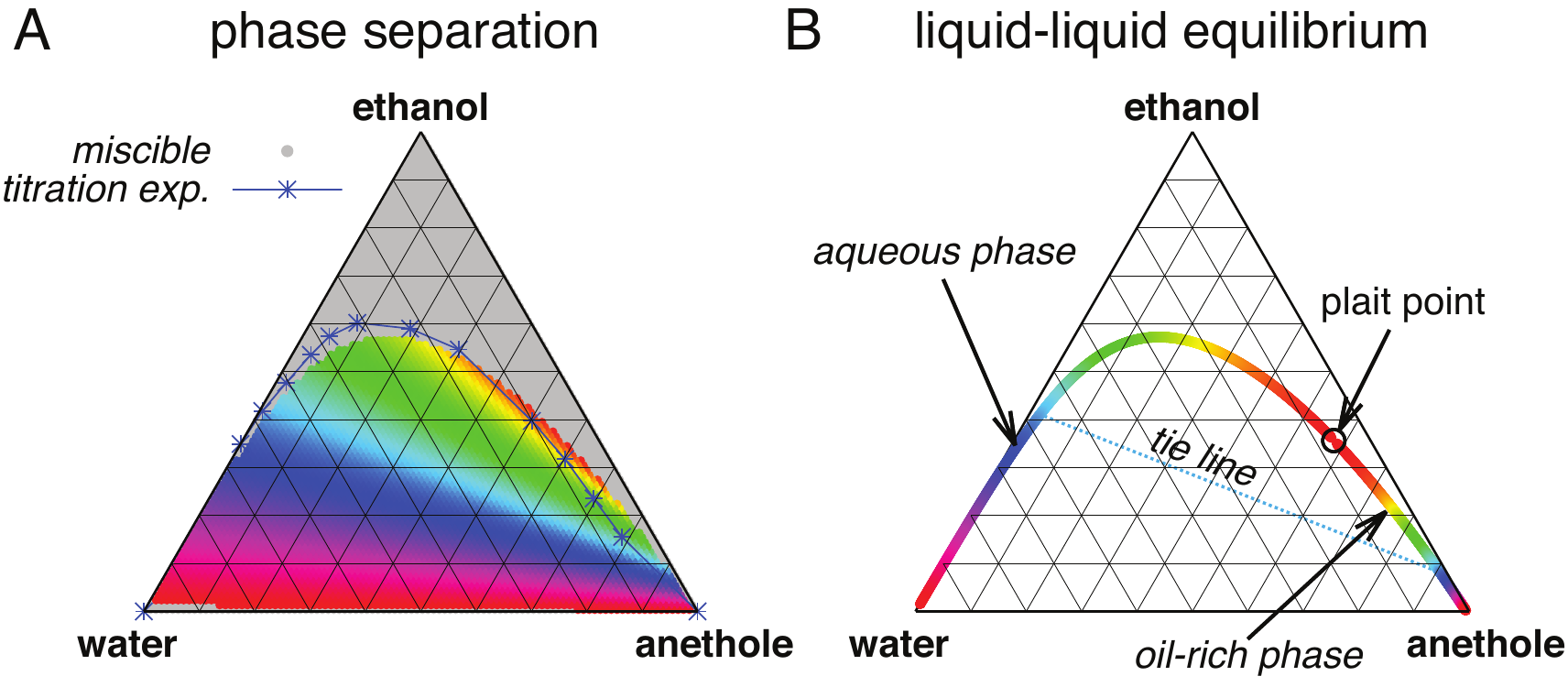}}
\caption{Phase separation predicted by UNIFAC. (A) Gray regions indicated homogeneous mixing, whereas phase separation is expected in the coloured region. The blue line with the blue stars (measured data points) indicates the good agreement with the titration experiments of \citet{tanouzo2017}. If the liquid is undergoing phase separation, the composition of the resulting two phases can be read off from the binodal in (B) by the color code.
As an example, we give the tie line for the turkish blue region.}
\label{fig:comp}
\end{figure}

The mixture can only undergo a phase separation if the mixture is nonideal, i.e. $\gamma_\alpha\neq 1$. Thus, the activity coefficients are fundamental to accurately describe the liquid-liquid equilibrium. Unfortunately, we have neither found sufficiently detailed experimental data for the present mixture, nor parameters for the UNIQUAC model, which was used by \citet{Chu2016}. Therefore, we employed the UNIFAC model \citep{fredenslund1975group}, which is more general than the UNIQUAC model, to perform the splitting of the molecules into functional subgroups. For detailed information about this model, we refer to \citet{Fredenslund1977}. Due to the better agreement with the titration experiments by \citet{tanouzo2017}, we took the recent modified UNIFAC (Dortmund) parametrization \citep{Constantinescu2016}. 

In order to find the liquid-liquid equilibrium curve and the regions of phase separation, we followed the method propsed by \citet{Zuend2010}: if equations \eqref{eqn:cp} can only be solved trivially, i.e. $x_\alpha^{\text{d}}=x_\alpha^{\text{h}}$, or if the Gibbs free energy at the trivial solution is below the Gibbs free energy of all non-trivial solutions, the mixture remains in a well-mixed single phase configuration. On the contrary, if there is a non-trivial solution with a lower Gibbs free energy than the trivial solution, phase separation is expected and the system will take on the solution with the globally minimal Gibbs free energy.

The data obtained by this procedure is depicted in Figure \ref{fig:comp}. In Figure \ref{fig:comp}A, the region of phase separation is shown in a ternary diagram in terms of mass fractions $m_{\alpha}$. In grey regions, the liquid at the specific composition remains perfectly mixed, whereas in the coloured regions, phase separation occurs. The composition of the two resulting phases can be read off from Figure \ref{fig:comp}B by the color code: The system splits into an aqueous phase and an oil-rich phase whose compositions are indicated by the respeative same color. Although the UNIFAC model is in general not perfect due to its dependence on the parameter table, the comparison of the phase separation region with the titration data of \citet{tanouzo2016} in Figure \ref{fig:comp}A (blue stars) show very good agreement.

\subsection{Mass transport}
\label{sec:mass}
\begin{figure}
\centering{
\includegraphics[width=0.7\textwidth]{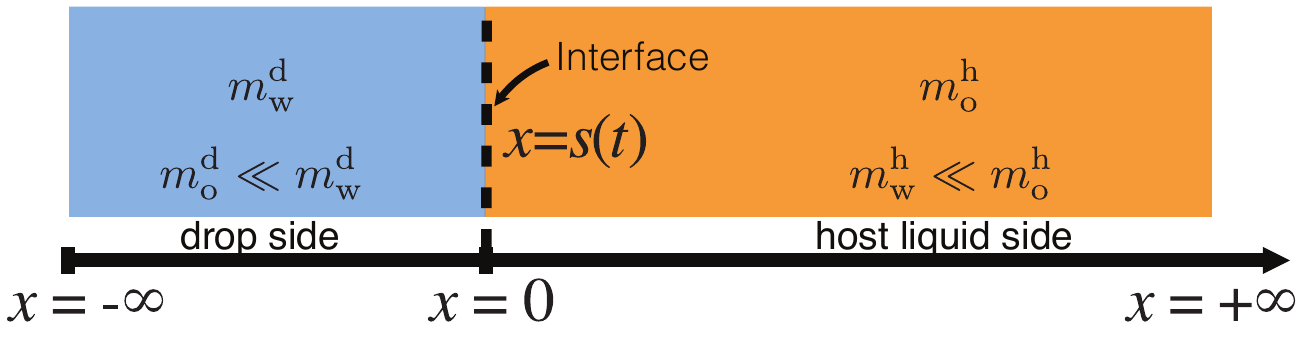}}
\caption{Sketch of the one-dimensional multiphase and multicomponent diffusion model with a moving interface at $x=s(t)$ separating the two regions of drop (d) and host liquid (h).
The subscripts stand for the species water (w) and anethole oil (o).
The mass fractions of the different species in the two different regions $m_{\alpha}^{\text{d}}$ and $m_{\alpha}^{\text{h}}$ are function of time $t$ and position $x$. The initial position of the interface $s(0)$ is defined to be at the origin.}
\label{fig:model}
\end{figure}

The system under discussion is composed of three species in two different phases, i.e. it is a multicomponent and multiphase diffusion system.
The diffusion process is modelled in a one-dimensional infinite space, as illustrated in Figure \ref{fig:model}.
The diffusion process of each individual constituent is assumed to depend only on its own concentration gradient in radial direction (Fick's diffusion laws).
$m_{\alpha}$ denotes the mass fraction of the species water ($\alpha=\text{w}$), ethanol ($\alpha=\text{e}$) and anethole oil ($\alpha=\text{o}$) as function of time $t$ and position $x$.
Again the superscripts indicate the region, i.e. either drop region (d) or host liquid region (h). The two regions are separated by the moving interface at position $s(t)$.
The initial position of the interface is set to the origin, i.e. $s(0)=0$.
The mass transport of each component $\alpha$ is governed by diffusion equations, namely
\begin{subequations}
\begin{align}
\frac{\partial m_{\alpha}^{\text{d}}}{\partial t} &= D_{\alpha}^{\text{d}} \frac{\partial^2 m_{\alpha}^{\text{d}}}{\partial x^2}, &-\infty<  x \leq s(t),\\
\frac{\partial m_{\alpha}^{\text{h}}}{\partial t} &= D_{\alpha}^{\text{h}} \frac{\partial^2 m_{\alpha}^{\text{h}}}{\partial x^2},  & s(t)\leq x< +\infty,\\
\big(m_{\alpha}^{\text{h}}-m_{\alpha}^{\text{d}}\big)\frac{\diff s}{\diff t} &= D_{\alpha}^{\text{d}} \frac{\partial m_{\alpha}^{\text{d}}}{\partial x}-D_{\alpha}^{\text{h}} \frac{\partial m_{\alpha}^{\text{h}}}{\partial x},    &\text{when}\quad x=s(t),
\end{align}
\label{eqn:ge}%
\end{subequations}
where the first equation describes the diffusion process in the drop region with a corresponding diffusivity $D_\alpha^{\text{d}}$, and the second one for the host liquid with diffusivity $D_\alpha^{\text{h}}$.
The Stefan condition is considered in the third equation, implying mass balances at the interface between the change rate of local mass due to the interface movement (lhs) and the combined diffusive mass flux of the species at the interface (rhs).
The composition in both phases is subject to the constraints
\begin{subequations}
\begin{align}
m_{\text{w}}^{\text{d}}+m_{\text{e}}^{\text{d}}+m_{\text{o}}^{\text{d}}=1,\\
m_{\text{w}}^{\text{h}}+m_{\text{e}}^{\text{h}}+m_{\text{o}}^{\text{h}}=1.
\end{align}
\label{eqn:me}%
\end{subequations}
Therefore, only two of three species, for instance water ($\alpha=\text{w}$) and anethole oil ($\alpha=\text{o}$), are solved by equations (\ref{eqn:ge}), and the third one, ethanol ($\alpha=\text{e}$), is calculated by this relationship.

As we only focus on the early stage of the diffusion process, the boundaries at $\pm\infty$ do not influence the dynamics in the vicinity of the interface. 
Therefore, the initial condition and the far field boundary condtitions are given by
\begin{subequations}
\begin{align}
m_{\alpha}^{\text{d}} &= m_{\alpha0}^{\text{d}}, &-\infty<  x < 0,\, t=0,\\
m_{\alpha}^{\text{h}} &= m_{\alpha0}^{\text{h}},  &0<  x < \infty, \, t=0,\\
m_{\alpha}^{\text{d}} &= m_{\alpha0}^{\text{d}}, &x=-\infty, \, t>0,\\
m_{\alpha}^{\text{h}}&= m_{\alpha0}^{\text{h}}, & x=\infty, \, t>0.
\end{align}
\label{eqn:icbc}%
\end{subequations}
The initial values are input from experimental conditions. 
This is possible as we know the initial composition of the drop mixture and the host liquid.
Denoting the interfacial compositions at the drop and the host liquid side of the interface as $m_{\alpha s}^{\text{d}}$ and $m_{\alpha s}^{\text{h}}$, respectively, we have
\begin{subequations}
\begin{align}
m_{\alpha}^{\text{d}}&= m_{\alpha s}^{\text{d}}, & x&=s(t),\,  t>0\\
m_{\alpha}^{\text{h}} &= m_{\alpha s}^{\text{h}}, & x&=s(t),\,  t>0.
\end{align}
\label{eqn:sc}%
\end{subequations}
The interfacial composition on the two sides of the interface are calculated by applying the liquid-liquid equilibrium constraint at the interface condition, cf. Sec. \ref{sec:LLE}.
In a ternary system, the interfacial composition has four degrees of freedom by virtue of mass conservation of the mixture.
The composition on two sides of the interface are situated on the binodal and are connected by a tie line in the phase diagram of the ternary system.
Therefore, in accordance with Gibbs' phase rule, the number of degrees of freedom for the interfacial composition reduces to one (for instance $m_{\text{w}s}^{\text{d}}$).

The solution of the governing diffusion equations (\ref{eqn:ge}) with the given conditions (\ref{eqn:icbc}) reads
\begin{subequations}
\begin{align}
m_{\alpha}^{\text{d}}=\mathcal{B}_{\alpha}^1+\mathcal{B}_{\alpha}^2\,\erf \frac{x}{2\sqrt{D_{\alpha}^{\text{d}}t}},\\
m_{\alpha}^{\text{h}}=\mathcal{B}_{\alpha}^3+\mathcal{B}_{\alpha}^4\,\erf \frac{x}{2\sqrt{D_{\alpha}^{\text{h}}t}},
\end{align}
\label{eqn:form}%
\end{subequations}
where $\mathcal{B}_{\alpha}^1$, $\mathcal{B}_{\alpha}^2$, $\mathcal{B}_{\alpha}^3$ and $\mathcal{B}_{\alpha}^4$ are coefficients to be determined.
By applying the additional conditions (\ref{eqn:sc}) to the solution form, we obtain them as
\begin{subequations}
\begin{align}
\mathcal{B}_\alpha^1&=\frac{m_{\alpha 0}^{\text{d}}\, \erf \frac{s}{2\sqrt{D_{\alpha}^{\text{d}}t}}+m_{\alpha s}^{\text{d}}}
{\erf \frac{s}{2\sqrt{D_{\alpha}^{\text{d}}t}}+1},\\
\mathcal{B}_\alpha^2&=\frac{m_{\alpha s}^{\text{d}}-m_{\alpha 0}^{\text{d}}}
{\erf \frac{s}{2\sqrt{D_{\alpha}^{\text{d}}t}}+1},\\
\mathcal{B}_\alpha^3&=\frac{m_{\alpha 0}^{\text{h}}\, \erf \frac{s}{2\sqrt{D_{\alpha}^{\text{h}}t}}-m_{\alpha s}^{\text{h}}}
{\erf \frac{s}{2\sqrt{D_{\alpha}^{\text{h}}t}}-1},\\
\mathcal{B}_\alpha^4&=\frac{m_{\alpha s}^{\text{h}}-m_{\alpha 0}^{\text{h}}}
{\erf \frac{s}{2\sqrt{D_{\alpha}^{\text{h}}t}}-1},
\end{align}
\label{eqn:coe}%
\end{subequations}
i.e., they seem to depend on $s(t)/\sqrt{t}$ only.
However, following the way successfully used by \citet{crank1987free}, \citet{kirkaldy1963}, and \citet{kenneth1972}, we assume the movement to be proportional to $\sqrt{t}$, which is consistent with the fact that the interface movement is driven by diffusion, i.e.
\begin{equation}
s(t)=\lambda \sqrt{t}, \qquad \text{and} \, \sqrt{D_s}=|\lambda|,
\label{eqn:s}
\end{equation}
where the sign of the constant $\lambda$ gives the direction of the interface movement, and $D_s$ can be regarded as the ``diffusion coefficient" of the moving interface.
The time-dependence cancels out, i.e., the coefficients $\mathcal{B}_\alpha^{i}$ are time-independent and only depend on $\lambda$.

To simplify the expression of the equations, we introduce the definitions
\begin{equation}
\mathcal{K}_\alpha=\frac{\lambda}{2\sqrt{D_\alpha^\text{h}}}, \quad 
\mathcal{R}_\alpha=\sqrt{\frac{D_\alpha^\text{h}}{D_\alpha^\text{d}}}, \quad
\chi_{\text{o}}^{\text{w}}=\sqrt{\frac{D_\text{o}^\text{h}}{D_\text{w}^\text{h}}},
\label{eqn:co}
\end{equation}
which together with equation (\ref{eqn:s}) are substituted into the coefficients (\ref{eqn:coe}) and the Stefan condition (\ref{eqn:ge}c).
Thereby, we obtain the coefficient constants as
\begin{subequations}
\begin{align}
\mathcal{B}_\alpha^1&=\frac{m_{\alpha 0}^{\text{d}}\, \erf (\mathcal{R}_\alpha \mathcal{K}_\alpha)+m_{\alpha s}^{\text{d}}}
{\erf (\mathcal{R}_\alpha \mathcal{K}_\alpha)+1},\\
\mathcal{B}_\alpha^2&=\frac{m_{\alpha s}^{\text{d}}-m_{\alpha 0}^{\text{d}}}
{\erf (\mathcal{R}_\alpha \mathcal{K}_\alpha)+1},\\
\mathcal{B}_\alpha^3&=\frac{m_{\alpha 0}^{\text{h}}\, \erf \mathcal{K}_\alpha-m_{\alpha s}^{\text{h}}}
{\erf \mathcal{K}_\alpha-1},\\
\mathcal{B}_\alpha^4&=\frac{m_{\alpha s}^{\text{h}}-m_{\alpha 0}^{\text{h}}}
{\erf \mathcal{K}_\alpha-1},
\end{align}
\label{eqn:b}
\end{subequations}
as well as the new expression of mass conservation (\ref{eqn:ge}c)
\begin{equation}
\sqrt{\pi}\mathcal{K}_\alpha \big(m_{\alpha s}^{\text{h}}-m_{\alpha s}^{\text{d}}\big)=
4 e^{-\mathcal{K}_\alpha^2}
\big[-\mathcal{B}_\alpha^4+e^{\mathcal{K}_\alpha^2(1-\mathcal{R}_\alpha^2)}\mathcal{B}_\alpha^2/\mathcal{R}_\alpha \big].
\label{eqn:sc-n}
\end{equation}

We thus have a closed-form solution for the problem, given by the mass balance equation (\ref{eqn:sc-n}) in adjunction with thermodynamic equilibria theory.
The number of the mass balance equations is two, one for water ($\alpha = \text{w}$) and the other for anethole oil ($\alpha=\text{o}$) and the five unknown variables are $m_{\text{w}s}^{\text{d}}$, $m_{\text{o}s}^{\text{d}}$, $m_{\text{w}s}^{\text{h}}$, $m_{\text{o}s}^{\text{h}}$, and $\lambda$.
As discussed above, the composition on two sides of the interface are situated on the binodal and are connected by a tie line in the phase diagram, see Figure \ref{fig:comp}B. Due to these equilibrium constraints, the number of degrees of freedom of the interfacial composition is one (for instance $m_{\text{w}s}^{\text{d}}$).
Hence, there are in total two unknown variables, for instance $m_{\text{w}s}^{\text{d}}$ and $\lambda$, in the equation.
By applying Newton's method with a given initial guess, we can find roots to the two equations (\ref{eqn:sc-n}) (one for $\alpha=\text{w}$ and one for $\alpha=\text{o}$) and the whole problem is solved.
All quantities defined in the model are list in Table \ref{tab:listall} given in the appendix \ref{sec:B}.

\subsection{Diffusion coefficients}
\label{sec:Dcoe}
In the model, there are in total four different diffusion coefficients, namely in each of the two different regions one coefficient for one of the two species: water diffusivity in the drop medium $D_{\text{w}}^{\text{d}}$, anethole oil diffusivity in the drop medium $D_{\text{o}}^{\text{d}}$, water diffusivity in the host liquid medium $D_{\text{w}}^{\text{h}}$, and anethole oil diffusivity in the host liquid medium $D_{\text{o}}^{\text{h}}$.

To acquire the diffusivities, we were confined to models and assumptions, since the direct measurement is complicated.
In the drop medium at the early stage,  the trans-anethole content is negligible, so that the coefficient $D_{\text{w}}^{\text{d}}$ is assumed to be given by the mutual diffusivity in a binary water-ethanol mixture. The corresponding values were obtained by fitting experimental data from \citet{Parez2013a}.
The coefficient $D_{\text{o}}^{\text{d}}$, i.e. dilute anethole in a water/ethanol mixture, is calculated in the limit of infinite dilution based on the model of \citet{Perkins1969}, and is listed in Appendix \ref{sec:A4}.
Both $D_{\text{w}}^{\text{d}}$ and $D_{\text{o}}^{\text{d}}$ depend on the initial ethanol content in drop.   

The model of \citet{Hayduk1982} was used to calculate the coefficients of dilute water in host anethole oil $D_{\text{w}}^{\text{h}}$.
The corresponding diffusivity constant $D_{\text{o}}^{\text{h}}$ is estimated according to the Stokes-Einstein equation, which states that the diffusion coefficient is inversely proportional to the size of molecules at constant viscosity. 
Since both ethanol and water content in the host liquid are negligible, we take $D_{\text{w}}^{\text{h}}=\SI{1.2e-9}{\meter^2\per\second}$ and $D_{\text{o}}^{\text{h}}=\SI{0.66e-9}{\meter^2\per\second}$, both of which are assumed to be constant.

\section{Model predictions}
\label{sec:prediction}
\subsection{Concentration profiles}

\begin{figure}
\centering{
\includegraphics[width=\textwidth]{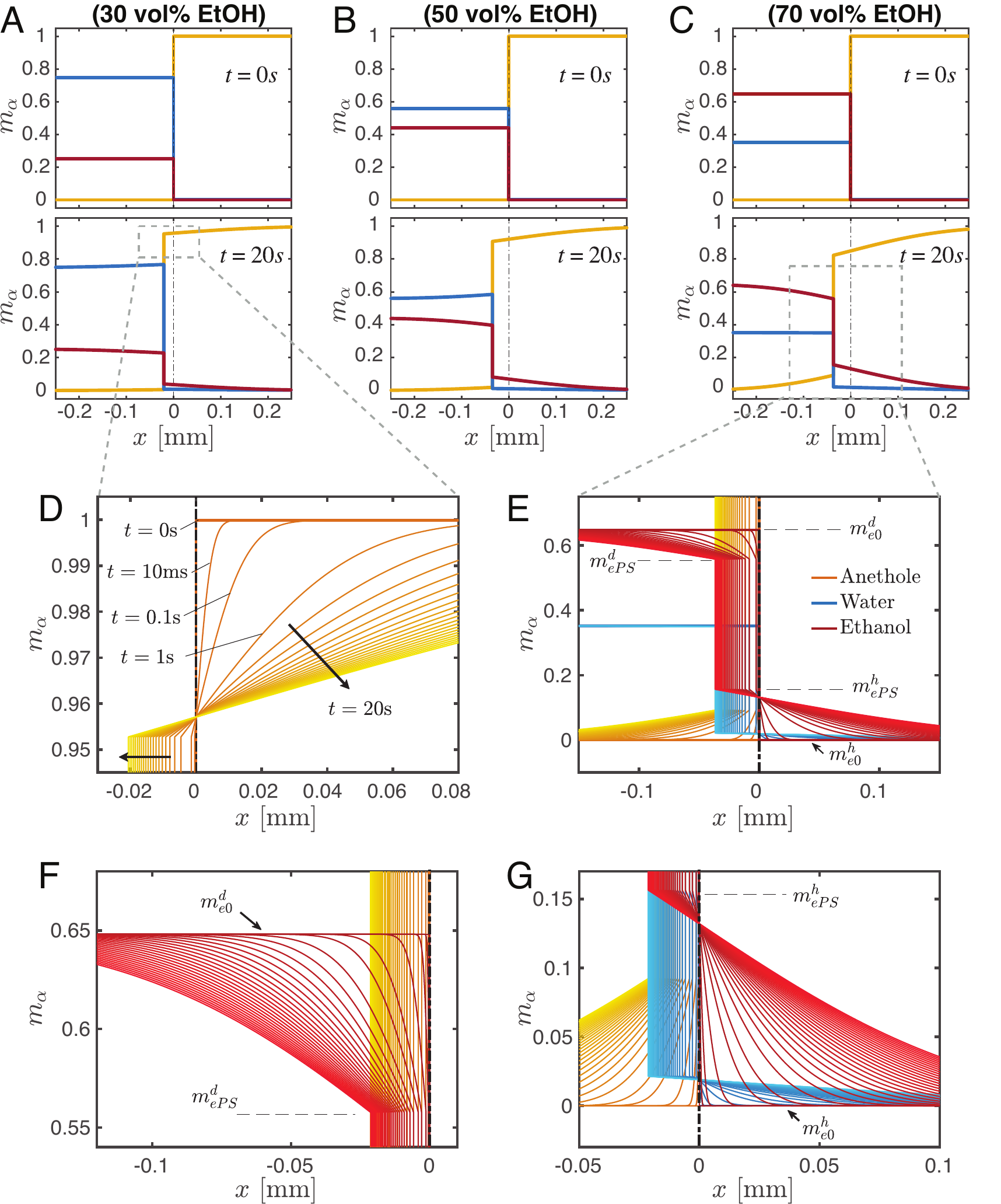}}
\caption{Calculated results of the diffusion model using the parameters described in the text.
(A-C) Snapshots of the mass fraction profile at \SI{0}{\second} (first row) and \SI{20}{\second} (second row) for different cases.  The vertical black dash-dotted line indicates the initial position of the interface, which separates the drop solution on the left side of figures and the host liquid on the right side. 
Note that the $m_{\alpha}$ represent mass fractions, which is different from the volume fractions.
Panels D and E are zoom-in pictures of panels A and C at $t=\SI{20}{\second}$, showing the profile evolution. 
Panels F and G are the zoom-ins of panel E.}
\label{fig:m}
\end{figure}

The mass concentration distribution of each constituent as function of time is obtained from the model described above.
Figure \ref{fig:m}ABC and corresponding close-up images, Figure \ref{fig:m}DEFG, show the development of the concentration profiles for different cases of binary drops with different initial water-ethanol ratios (70:30 v/v in A, 50:50 v/v in B, and 30:70 v/v in C, animations of 40:60 v/v and 60:40 v/v are available \textcolor{blue}{in supplementary material Movie S5}).
Anethole oil, water, and ethanol are labeled with yellow, blue, and red colours, respectively.
The vertical black dashed line stands for the initial position of the interface between the aqueous water-ethanol region (left side) and the anethole oil-rich region (right side).
Initially, the concentration profile of each species is a step function (first row of Fig. \ref{fig:m}ABC), which is consistent with the initial condition (\ref{eqn:icbc}ab) of the model.
Different initial ethanol content in the drop accounts for the difference of the step heights in the figures.

In the close-up Figure \ref{fig:m}D, once the diffusion starts, the discontinuity in the profile at $t=\SI{0}{\second}$ is immediately smoothed with a new turning point emerging. 
The dynamics of the concentration profiles is apparent from the temporal curves at $t=\SI{10}{\milli\second}$, $t=\SI{1}{\second}$, $t=\SI{2}{\second}$ and so on.

The concentration gradient at the interface induces a diffusive flux for each constituent towards or from infinity (anethole in Fig. \ref{fig:m}D, ethanol and water in Fig. \ref{fig:m}E). The step height at infinity remains unchanged due to the Dirichlet boundary condition (\ref{eqn:icbc}cd).
In the employed Stefan problem, the interface has the freedom to shift towards the aqueous region (left side), for all the constituents in the system. 
After \SI{20}{\second}, the developed concentration profiles in different cases are displayed in the second row of Figure \ref{fig:m}ABC. 
The height of the new turning point, corresponding to the interfacial concentration $m_{\alpha s}^{\text{d}}$ and $m_{\alpha s}^{\text{h}}$ (ethanol, $\alpha=\text{e}$, in Fig. \ref{fig:m}E), is fixed during the movement of the interface, which reflects the assumption of the stability of liquid-liquid equilibrium at interface.
Figures \ref{fig:PDG}F and G show further zoom-ins of the interface.

\subsection{Diffusion paths theory and calculated results}
The diffusion path method was proposed by \citet{kirkaldy1963}, who mapped the composition of the ternary solution on its phase diagram (Fig.\ref{fig:comp}B).
The values of the composition along the domains are mapped in the diagram as a line, known as \textit{diffusion path} or composition path.
It provides an effective way of representing the relationship between kinetic and thermodynamic aspects for a multiphase and multicomponent system.

In the spirit of the pioneering work by \citet{kenneth1972}, we predict the diffusion-induced spontaneous emulsification by examining the geometrical relationship between diffusion paths and binodal curve.
If the diffusion path crosses the binodal curve (Fig.\ref{fig:comp}B) between start and end point, supersaturation in the media is present, which induces spontaneous emulsification.
It is noteworthy that the intersections of all considered diffusion paths with the binodal curve are far away from the plait point and the intersections are located at the edge of the ouzo effect region. 

In ternary system, the path is determined by any two independent constituents.
Since we know the evolution of the distribution of all species, we can mathematically express their relationship.
We know that the coefficients (\ref{eqn:b}ab) are constant for a certain case, which implies that equation (\ref{eqn:form}a) is well-determined.
Therefore, equation (\ref{eqn:form}a) gives functional relationships $m_{\text{w}}^{\text{d}}=f_{1}(\frac{x}{\sqrt{t}})$ and $m_{\text{o}}^{\text{d}}=f_{2}(\frac{x}{\sqrt{t}})$.
Upon substituting them into equation (\ref{eqn:me}a), we obtain the mathematical expression of the diffusion path in the drop region,
\begin{subequations}
\begin{align}
m_{\text{e}}^{\text{d}}=1-\mathcal{F}(m_{\text{o}}^{\text{d}})-m_{\text{o}}^{\text{d}},
\end{align}
\label{eqn:dp1}
\end{subequations}
where  $\mathcal{F}=f_{1}\cdot f_{2}^{-1}$.
By the same way, we obtain the expression of the diffusion path in the host liquid medium,
\begin{subequations}
\begin{align}
m_{\text{e}}^{\text{h}}=1-m_{\text{w}}^{\text{h}} -\mathcal{G}(m_{\text{w}}^{\text{h}}),
\end{align}
\label{eqn:dp2}
\end{subequations}
where $\mathcal{G}=g_{2}\cdot g_{1}^{-1}$,
given $m_{\text{w}}^{\text{h}}=g_{1}(\frac{x}{\sqrt{t}})$ and $m_{\text{o}}^{\text{h}}=g_{2}(\frac{x}{\sqrt{t}})$. 
Notably, the mathematical expressions reveal that the diffusion path is \textit{independent} of time and space.
In other words, the set of the compositions at a certain position for different moments and the set for a certain moment among different positions share the same diffusion path.
The composition at the end point of the diffusion path lies on the binodal curve and corresponds to the interfacial composition, whereas the composition at the start point of the diffusion path indicates the composition at infinity, which is the same as the initial composition in the domain.

\begin{figure}
\centering{
\includegraphics[width=\textwidth]{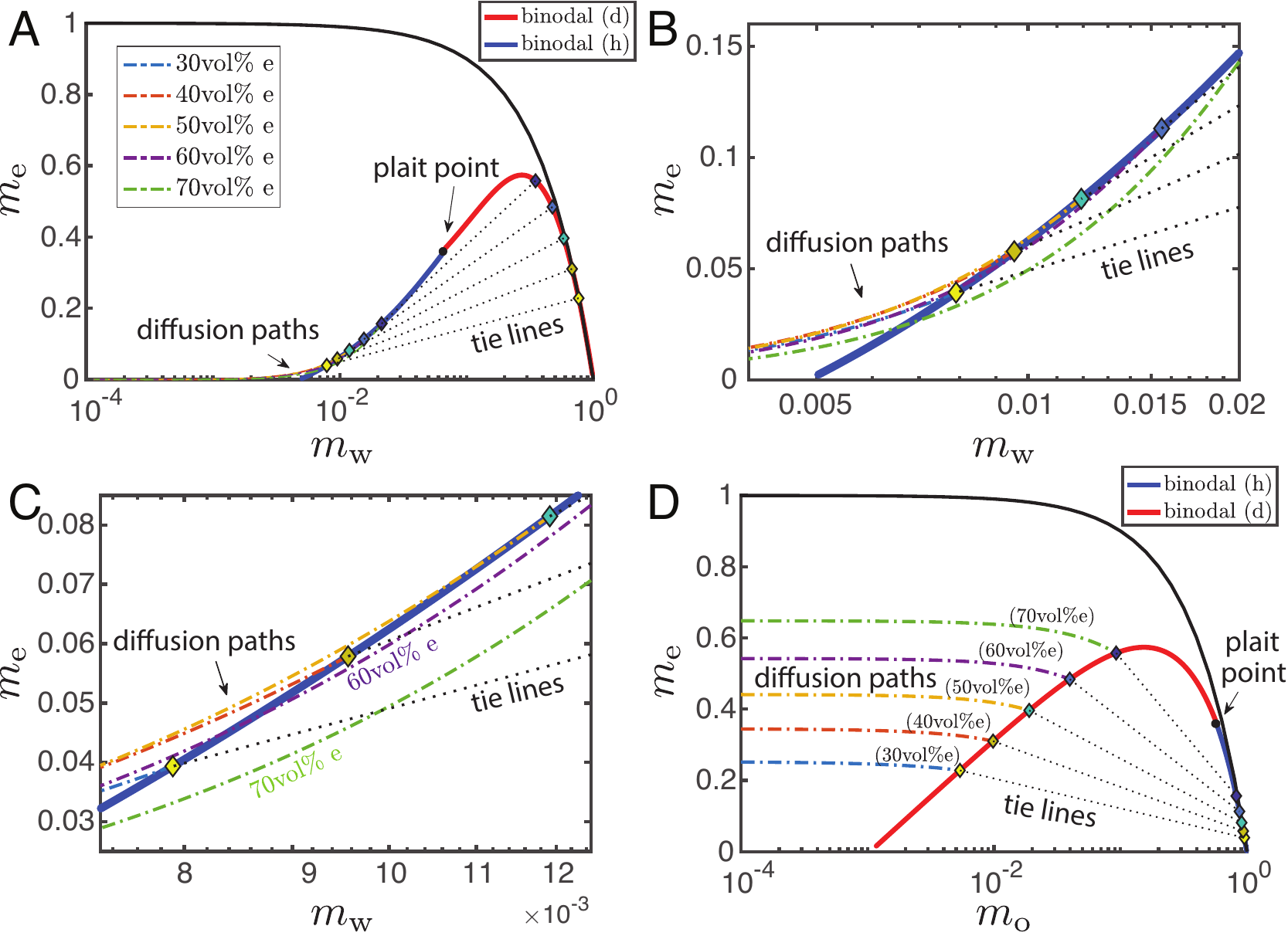}}
\caption{Phase diagrams showing the calculated diffusion paths in the water-ethanol phase diagram (A) and in the anethole-ethanol phase diagram (D).
The oil-rich part of binodal curve is labeled in blue and the water-rich part in red.
The black dotted lines are the tie lines.
Panels B and C are zoom-ins of panel A.
The close-up figures B and C show that the diffusion path in the host liquid passes through the binodal for the cases with high ethanol content ($\geq\SI{50}{vol\percent}$).
(D) In contrast, in the aqueous phase, in all cases, there is no diffusion path crossing the binodal curve.}
\label{fig:PDG}
\end{figure}

Figure \ref{fig:PDG} shows the calculated results, displayed in mass fraction phase diagrams of two independent constituents.
Figure \ref{fig:PDG}A with zoom-ins Figures \ref{fig:PDG}B and C shows the water-ethanol phase diagram, enhancing the oil-rich regime with w-in-o emulsions, whereas Figure \ref{fig:PDG}D shows the anethole-ethanol phase diagram, highlighting the water-rich regime with o-in-w emulsions.
The binodal curve is divided by the plait point into two different colours, blue for oil-rich phase and red for water-rich phase.
The tie lines, connecting the composition points belonging to both sides of the liquid-liquid equilibrium, were calculated by the method described in Section \ref{sec:LLE}.
Dot-dashed lines with different colours are the diffusion paths for different cases (same colour labeling for Fig. \ref{fig:PDG}ABCD). 
The close-up image in Figures \ref{fig:PDG}B and C reveal that, when the ethanol volume fraction in the aqueous phase is higher than \SI{51.98}{vol\percent}, corresponding to a transition mass fraction $m_{\text{e}*}^{\text{d}}$ calculated by the model, the diffusion path has to pass through the binodal curve to meet the equilibrium points (diamond dots) at the binodal.
The segment of diffusion path that is below the binodal is in supersaturation condition, which indicates the appearance of w-in-o emulsions in oil-rich side (host liquid region).
Whereas in the water-rich side (Fig. \ref{fig:PDG}D), all the diffusion paths meet equilibrium points without intersecting the binodal curve.
Therefore, according to the model, there is no supersaturation region induced by pure diffusion processes, which predicts the absence of o-in-w emulsions in drop region.
The time independence implies self-emulsification immediately occurs once the diffusion process starts.

\subsection{Self-emulsification: comparison between model predictions and experimental observations}

\begin{figure}
\centering{
\includegraphics[width=\textwidth]{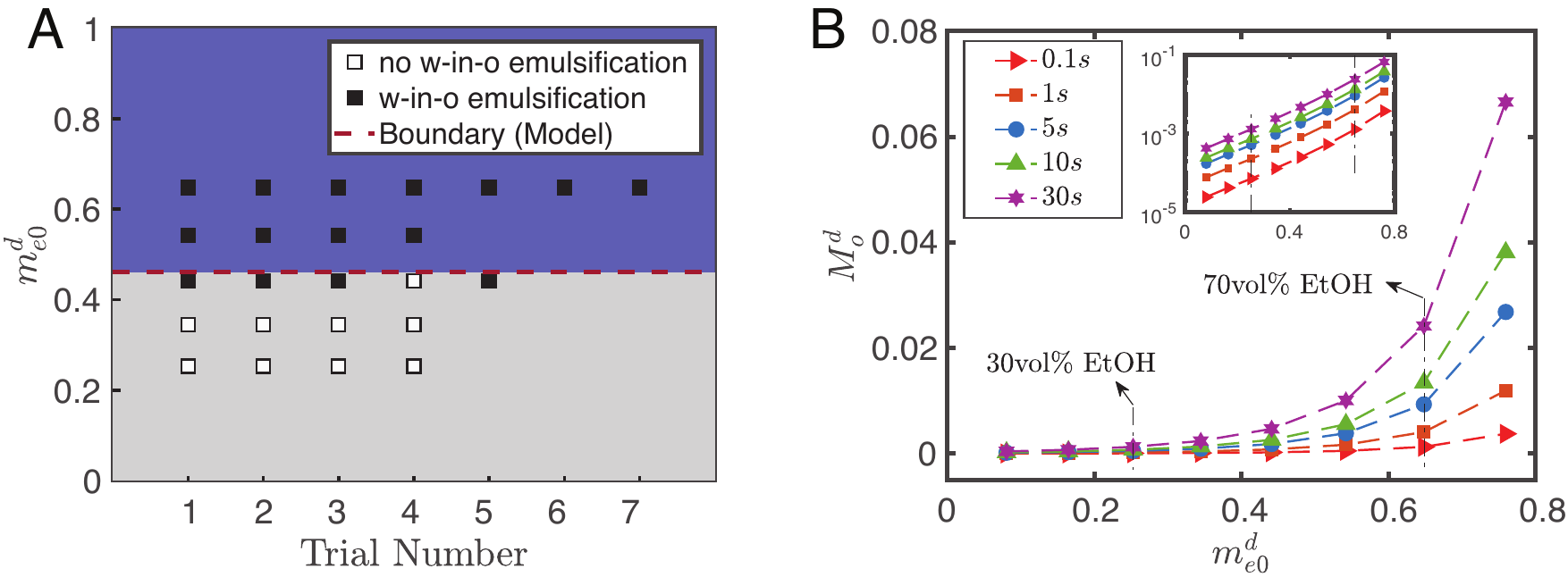}}
\caption{(A) Comparison between the model prediction and the experimental observation for the appearance of w-in-o emulsions outside the drop.
The ordinate is the initial mass fraction of ethanol in the drop and the abscissa gives the trial number.
The black and white dots are experimental observation data from Table \ref{tab:emu} and denote presence and absence of emulsification, respectively.
The red dashed line is a transition value calculated by the model, above which w-in-o emulsification is predicted to happen. 
(B) Calculation results of the total amount of oil transported into the drop as function of the initial ethanol content of the drop using the one-dimensional diffusion model.
Different symbols present the entered oil amount at different moments. The inserted figure gives the same data on a log-log plot.
}
\label{fig:PD}
\end{figure}

For w-in-o emulsion, the comparison between model prediction and experimental observation is presented in Figure \ref{fig:PD}A.
The vertical ordinate denotes the initial ethanol mass fraction in the water/ethanol drop and the abscissa gives the trial number, i.e. the individual experiments at this particular initial composition.
The data points are from Table \ref{tab:emu}: black symbols indicate the presence of w-in-o emulsion; white symbol means that the w-in-o emulsion is absent.
The red dashed line is the calculated transition percentage $m_{\text{e}*}^{\text{d}}$, above which (blue region) the model predicts the presence of w-in-o emulsion in the host liquid, otherwise absence (grey region).
The comparison reveals the reasonable predictive power.
In the blue region, all data points are black, whereas in the grey region far from transition line, all the data dots are white.
Around the transition line, both black and white data points exist.

For the o-in-w emulsion in the drop medium, no diffusion-tiggered self-emulsification was observed in experiments within half a minute (Tab. \ref{tab:emu}), which is consistent with the model prediction.
However, oil microdroplets must appear since the ethanol content in the drop is reducing over time, while the oil content is increasing.
At some point, the composition in the drop region will be supersaturated and cause the ouzo effect inside the drop. 
However, due to the infinite domain in the model with Dirichlet boundary conditions, the model cannot account for the long time behavior.
The onset time for the oil microdroplets is at least more than half a minute, and has a negative correlation with initial content of ethanol in the drop as recorded in Table\ref{tab:emu}.

Although we are unable to obtain the overall composition of the three-dimensional drop from the one-dimensional model, it is possible to gain more understanding on the transport of anethole oil during the early stage by calculating the total amount of anethole transported into drop $M_{\text{o}}^{\text{d}}$. 
Its definition is given by integrating the mass fraction of anethole $m_{\text{o}}^{\text{d}}$ on the interval of aqueous region from $-\infty$ to $s$,
\begin{equation}
M_{\text{o}}^{\text{d}}=\int_{-\infty}^{s} (m_{\text{o}}^{\text{d}} - m_{\text{o}0}^{\text{d}}) \diff x.
\label{eqn:M1}
\end{equation}
Then we have
\begin{equation}
M_{\text{o}}^{\text{d}}=\big( m_{\text{o} s}^{\text{d}} -m_{\text{o} 0}^{\text{d}}  \big)\,s
+ \frac{\big(s m_{\text{o} s}^{\text{d}}-m_{\text{o} 0}^{\text{d}}\big)s}{\sqrt{\pi}\mathcal{R}_\text{o} \mathcal{K}_\text{o}} \,
\frac{e^{-\mathcal{R}_\text{o}^2 \mathcal{K}_\text{o}^2}}
{\erf(\mathcal{R}_\text{o} \mathcal{K}_\text{o})+1}.
\label{eqn:M2}
\end{equation}
Figure \ref{fig:PD}B shows that a higher ethanol percentage in drop $m_{\text{e}0}^\text{d}$ leads to a higher rate of transport of oil into the drop $M_\text{o}^\text{d}$.
The inserted figure reveals that it is an exponential growth relationship.
So the difference in the oil amount in the drop caused by the ethanol content increases over time (from red solid $\rhd$ dot line, at \SI{0.1}{\second}, to purple solid $\davidsstar$ dot line, at \SI{30}{\second}).
This gives a possible explanation for the experimental observation, as a higher ethanol concentration in the drop leads to more oil after the early stage of the diffusion process, which favours the occurrence of the ouzo effect inside the drop.

\subsection{Model limitations}
Although the model provides information on the mass transport of the multiphase and multicomponent diffusion process, as well as a good prediction to the behaviour of spontaneous emulsification, it is subject to the following limitation:
(i) The diffusion model is developed as a one-dimensional model without consideration of flow motion, i.e. a pure multiphase and multicomponent diffusion process;
(ii) The dependence of diffusivity on the local species concentration is disregarded and off-diagonal diffusion terms, i.e. as in the Maxwell-Stefan theory, are not considered; 
(iii) We apply Dirichlet boundary conditions at infinity, which implies that the model is only applicable in the early stage of the diffusion process.
In the long-time limit, the finite size of the regions, in particular of the drop, will become important, since there is not an infinite reservoir for the species. 
(iv) The detailed process of phase separation at interface is not considered, as we apply an instantaneous liquid-liquid equilibrium assumption;

The flow motion in the system has a big impact on the position where emulsification takes place.
A strong flow rate has the capacity to prevent emulsification by changing the local concentrations or by directly dissolving the nucleated emulsion.
Therefore it is necessary to have an investigation on the fluid dynamics of the system in Section \ref{sec:hydrodynamics}.

\section{Discussion on the fluid dynamics of the system}
\label{sec:hydrodynamics}

\subsection{Scaling analysis}
In the studied system, diffusion-induced advection, in the laminar flow regime, can \textit{in turn} play a significant role for the diffusion process, i.e.
diffusion and advection are highly coupled in the system.
The diffusion causes non-uniformities in the species distribution in both drop and host liquid media, which drives bulk flow motions (advection).
The flow reorganises the concentration field and consequently, in turn, affects the diffusion.
The generated advection mainly encompasses solutal Marangoni flow and buoyancy-driven flow (\citet{Erik2016}), i.e. solutal natural convection, whereas thermal effects are negligible compared to the two of them. 
As solutal Marangoni flow and buoyancy-driven flow have alternating dominance inside respective outside the drop their impacts on diffusion can be treated separately.
To confirm that only one of the two respective flows is dominant in the respective domain, we perform the following scaling analysis.

The characteristic velocity of the solutal Marangoni flow $U_{\text{M}}$ scales like
\begin{equation}
U_{\text{M}} \sim \Delta \gamma / \mu,
\end{equation}
where $\mu$ is the dynamic viscosity (Appendix \ref{sec:A5}), and $\Delta \gamma$ is the interfacial surface tension difference caused by compositional variations.
The characteristic velocity associated with the buoyancy-driven convection $U_\text{B}$ is obtained by balancing the viscous dissipation rate within the convection roll with the gaining rate of potential energy due to gravity \citep{tam2009}, i.e.
\begin{equation}
\int {\mu (\nabla u)^2} \diff V \sim \Delta \rho g U_{\text{B}} \ell^3,
\label{eqn:UM}
\end{equation}
where the left side scales as $\mu \ell (U_{\text{B}})^2$. Here $\ell$ is the characteristic length scale and $g$ is the gravitational constant. 
Hence, the characteristic velocity of the buoyancy-driven flow $U_{\text{B}}$ scales as
\begin{equation}
U_{\text{B}} \sim \Delta \rho g \ell^2 /\mu.
\label{eqn:UB}
\end{equation}
With equations (\ref{eqn:UB}) and (\ref{eqn:UM}), we can estimate the ratio of buoyancy-driven flow to Marangoni flow by
\begin{equation}
\frac{U_\text{B}}{U_\text{M}} \sim \frac{\Delta \rho g \ell^2}{\Delta \gamma}.
\end{equation}
We can note the ratio is a Bond number, which measures the importance of gravitational forces compared to interfacial surface tension. The Bond number is proportional to $\ell^{2}$, which indicates that relative importance of the two flow mechanisms has a strong dependence on the spatial scale of the domain.

We can estimate the ratio inside the drop by taking the drop radius $R\sim \SI{0.75}{\milli\meter}$ as the length scale $\ell$.
Since the interfacial surface tension is composition-dependent and varies during drop dissolution, we estimated the difference $\Delta \gamma \sim \SI{12.1}{\milli\newton\per\meter},$ as half of the interfacial surface tension between pure water and anethole oil \citep{tanouzo2017}.
The density difference is selected as the biggest density difference between water and ethanol, i.e. $\Delta \rho = \rho_{\text{w}}-\rho_\text{e}$ ($\sim \SI{211}{\kilogram \per \meter^3}$).
The estimation shows that $U_\text{B}\sim  U_\text{M}/10$, which implies that Marangoni flow inside the drop is prevailing.
Outside the drop, the length scale $\ell$ is selected as the host liquid depth, $\ell=\SI{7.5}{\milli\meter}$.
Then the estimate of the velocity ratio becomes $U_\text{B}\sim 10 U_\text{M}$, which indicates that buoyancy-driven flow is dominating outside the drop.


\subsection{Flow motions and its influence on spontaneous emulsification}
Taking the understanding of the flow motion in the system from the scaling analysis, we disregard the Marangoni flow in the following discussion about the hydrodynamics in the host liquid (large scale domain), whereas the buoyancy effect is neglected inside the drop (small scale domain).
The hydrodynamics outside and inside drop are discussed successively.

As water-ethanol drop dissolves, ethanol diffuses into the surrounding anethole oil.
Then, buoyancy starts to play a role, as ethanol is less dense than water and anethole oil, i.e. $\rho_\text{e}<\rho_\text{o}$ and $\rho_\text{e}<\rho_\text{w}$ ($\rho_\text{w}= \SI{998}{\kilogram \per \meter^3}, \rho_\text{e}=\SI{787}{\kilogram \per \meter^3}, $ and $\rho_\text{o}=\SI{988}{\kilogram \per \meter^3}$ at \SI{22}{\celsius}).
The surrounding ethanol-rich oil floats up in a form of the solute plume, causing an upwelling flow. Simultaneously, fresh oil liquid far from the drop replenishes to achieve mass continuity, as sketched in Figure \ref{fig:sk}A. 
So a convection flow outside the drop forms as a consequence, which is clearly observed through PIV measurements (Fig. \ref{fig:sk}B).
In the region next to the tropic of capricorn of the drop and not far from the drop surface, there is weak flow, enclosed by the yellow circle in the close-up image.
It indicates a subtle influence of the convection on the diffusion process in this region, whereas in other regions, the refresh oil brought by the relatively strong flow prevents the formation of a local supersaturation.
This is the reason that accounts for the appearance of w-in-o microdroplets in the certain position with the weak flow rate.

The generation of the buoyancy-driven convection affects the distribution of the diffusion rate along the drop surface.
The convection flow brushes away the diffused ethanol next to the drop surface and varies the concentration distribution of ethanol in the surrounding oil.
Around the equator of the drop the concentration boundary layer is thin, due to the intense inflow of bulk oil without ethanol (Fig. \ref{fig:sk}B): the normal ethanol concentration gradient outside the liquid-liquid interface $(\partial_{r}c^{\text{e}})_{\text{side}}$ has a steep slope.
At the top of the drop, the ethanol concentration gradient $(\partial_{r}c^{\text{e}})_{\text{top}}$ may also be increased, but only to a lower extent, as the replenishing oil is already contaminated with ethanol (and water) during its travel along the drop surface.
Due to the boundary condition, the replenishment of fresh oil caused by the convection is suppressed near the corner of the drop, where the ethanol concentration gradient $(\partial_{r}c^{\text{e}})_{\text{C.L.}}$ is much less affected.
Therefore, the buoyancy-driven convection predominantly enhances the diffusion flux near the drop equator rather than that above the drop or in the contact region and, as a consequence, the radial gradients of ethanol concentration along the drop surface obey the relationships
\begin{subequations}
\begin{align}
(\partial_{r}c^{\text{e}})_{\text{side}} &>(\partial_{r}c^{\text{e}})_{\text{top}} ,\\
(\partial_{r}c^{\text{e}})_{\text{side}} &>(\partial_{r}c^{\text{e}})_{\text{C.L.}} .
\end{align}
\label{eqn:cg}%
\end{subequations}
The two different gradients of the ethanol concentration result in a inhomogeneneous ethanol concentration along the inside of the interface.
The intense concentration gradient of ethanol around the equator $(\partial_{r}c^{\text{e}})_{\text{side}}$ generates a large ethanol diffusion flux, resulting in a locally higher water concentration inside the drop.
As the interfacial surface tension between oil phase and aqueous phase has a positive correlation with the water concentration in aqueous phase,
the interfacial surface tension has gradients from the south pole and north pole towards the equator.
Thus, two solute Marangoni convections arise following the gradients, as sketched in Figure \ref{fig:sk}A.
Meanwhile, the high water concentration around the equator of the drop accounts for the prefered emulsification there (Fig. \ref{fig:snap}C).

\section{Summary and conclusions}

We have experimentally presented the rich phenomena of water/ethanol drops dissolving in oil as host liquid, which encompass w-in-o emulsification outside the drop, o-in-w emulsification inside the drop, buoyancy-driven convection dominating outside the drop, and prevailing solutal Marangoni convection inside the drop.
O-in-w emulsification occurs around half minute later than w-in-o emulsification and w-in-o emulsification does not occur when reducing the initial ethanol concentration of the drop.

A quantitive understanding and the predictions of the diffusion-induced emulsification were theoretically achieved by developing a one-dimensional multiphase and multicomponent diffusion model, which incorporates thermodynamical equilibrium theory and diffusion path theory.
The prediction of the model agrees with experimental observations: diffusion-triggered w-in-o emulsification occurs when the drops have a ethanol content higher than \SI{51.98}{vol\percent}; o-in-w emulsification cannot be induced by pure diffusion.
Due to the infinite domain and the Dirichlet boundary conditions used in the model, the model is only applicable for the early stage of the multiphase diffusion process.
In practice, the continuous reduction of ethanol and increasing of oil in the drop lead to the occurrence of the o-in-w emulsification in a long time, which is thereby independent of the initial ethanol content of the drop.

A scale analysis and the experimental investigation of the diffusion-induced flow motion were performed to gain insight into its influence on the emulsification process. 
By the scale analysis, we demonstrated that in drop domain, the solute Marangoni flow prevails over the buoyancy-driven flow, while in host liquid domain, the latter dominates.
The buoyancy-driven convection enhances the ethanol diffusion rate around the equator of the drop, which gives rise to a reduction of the local ethanol concentration around the equator of the drop, arising of two Marangoni convection rolls inside the drop with opposite directions, and the generation of a preferred position for o-in-w emulsification. 
Due to the buoyancy-driven flow, fresh oil is replenished at the interface. 
Therefore w-in-o emulsification can only occur around a region next to the tropic of capricorn of the drop, where only weak flow and replenishment is present.

Although in this paper we provided a systematic study about the emulsification triggered by the dissolution of multicomponent drop in a host liquid, further investigations and discussions are required.
A more comprehensive model can be generated, either by considering the finite domain size and the appropriate boundary conditions, or by taking fluid motion into account and developing an axisymmetric model.
Experimentally, the  influence of temperature, different chemical systems, or different geometries by varying the contact angle of the drop are appealing open questions to explore.
A better understanding of the dissolution of multiphase and multicomponent systems may provide valuable information for the investigation of multiphase systems and spontaneous emulsifications in general. 
We also hope that our work may provide contribution to industrial application, such as modern liquid-liquid microextraction techniques.


\appendix
\section{Experimental details}
\subsection{Set-up and image analysis}
\label{sec:A1}
The experiments were performed in a lab without people around during the data recording. 
The temperature of the host liquid in the cuvette was around \SI{22}{\celsius}. 
The dissolution processes of the drops were observed by three synchronized cameras, one monochrome CCD camera (Ximea; MD061MU-SY) attached to a long-distance microscope system (Infinity; Model K2 DistaMax) for side view recordings, 
which was used for the drop profile detection, one digital SLR camera (Nikon; D750) equipped with a CMOS sensor attached to a high-magnification zoom lens system (Thorlabs; MVL12X3Z) for the other side view recordings, which was used for the side-view observation of the emulsification process, and another digital SLR camera (Nikon; D5100) equipped with a CMOS sensor attached to an identical high-magnification zoom lens system (Thorlabs; MVL12X3Z) for top view recordings, used for top-view observation of the emulsification process.
A cold light source (Olympus; ILP-1) was positioned at the same side as the SLR camera to illuminate the emulsion.

We performed image analyses with a custom-made MATLAB program.
The monochrome image series from the side view recordings were utilized in obtaining the temporal evolution of the dissolving characteristics of the drops.
For each image, we first pre-processed the data to increase the image contrast and then calculated the profile of the drop using the Canny method \citep{canny}.
The data points of the detected profile were fitted by part of a circle, i.e. assuming that the droplet is in a spherical cap shape.
The diameter of drop contact area $L$ and the contact angle $\theta$ were calculated based on the geometrical relationship between the base line of the surface and the fitted spherical cap. 
The drop volume $V$ was calculated by integrating the volumes of the horizontal disk layers.

\subsection{Particle image velocimetry}
\label{sec:A2}
We qualified the confidence level of the tracking particles following the flow by calculating the Stokes number and the ratio of the Stokes number to a buoyancy-corrected Froude number \citep{varghese2016}.
The relaxation time of the particle was estimated by
\begin{equation}
t_0\equiv (1+\rho_f/(2 \rho_p))\frac{\rho_p d_p^2}{18 \mu_f},
\label{eqn:relaxation}
\end{equation}
with the consideration of added mass force \citep{Oliveira2015}, where $\rho_f = \SI{0.988}{\gram\per\centi\meter^3}$ and $\mu_f = \SI{4.2}{\milli\pascal \cdot \second}$ are the density and the dynamic viscosity of the host liquid (trans-anethole), $\rho_p = \SI{1.03}{\gram\per\centi\meter^3}$ and $d_p = \SI{5}{\micro\meter}$ the density and the diameter of the tracking particles.
The Stokes number $St\equiv t_0 u_{\text{max}}/R$ can be calculated as
\begin{equation}
St= (1+\rho_f/(2 \rho_p))\frac{\rho_p u_{\text{max}} d_p^2}{18 \mu_f R}\sim10^{-5}\ll1,
\label{eqn:st}
\end{equation}
where $u_{max}$ is the maximum fluid velocity ($\sim \SI{10}{\milli\meter\per\second}$), $R$ the initial radius of the drop ($\sim \SI{0.5}{\milli\meter}$).
The buoyancy-corrected Froude number was defined as
\begin{equation}
Fr \equiv \frac{{u_{max}^2/R}}{(1-\rho_f/\rho_p)g},
\label{eqn:fr}
\end{equation}
taking the particle density, through $(1-\rho_f/\rho_p)$, into account. With eqn. (\ref{eqn:st}) and eqn. (\ref{eqn:fr}) we have
\begin{equation}
St/Fr \sim 0.3<1,
\label{eqn:frb}
\end{equation}
which combined with eqn. (\ref{eqn:st}) reveals that the tracking particles exactly follow the host liquid flow surrounding the dissolving drop \citep{varghese2016}.

\subsection{Detecting density differences}
\label{sec:A3}
They are visible thanks to the variation in light transmission when oil is mixed with ethanol.
The transmission has a negative correlation with the refractive index according to the Beer-Lambert law.
The refractive index of ethanol, \SI{1.361}{}, is smaller than that of trans-anethole, \SI{1.561}{}, and accounts for a better light transmission.
In our setup, the cold light source and side-view colour camera for the emulsion observations were positioned on the same side, which indicates that the material with a better light transmission is dimming.
Thus, it verifies that the solute plume is indeed an ethanol-rich oil mixture.
The variation in the refractive index also causes the distortion of the light path and leads to a locally hazy scene in the top views (Fig. \ref{fig:snap}ABCD).

\subsection{Diffusivity}
\label{sec:A4}
The diffusivity of dilute anethole in a water/ethanol mixture was calculated based on the model of \citet{Perkins1969}.

\begin{minipage}{\linewidth}
\def~{\hphantom{0}}
\begin{tabular}{l c c c c c}
\hline 
   w:e ratios & 3:7& 4:6 & 5:5 & 6:4 & 7:3 \\
    \hline
   $D_{\text{o}}^{\text{d}} \,[\SI{}{\meter^2\per\second}]$ & \SI{3.99e-10}{} & \SI{4.13e-10}{} & \SI{4.29e-10}{} & \SI{4.49e-10}{} & \SI{4.74e-10}{} \\
\hline
\end{tabular}
\label{tab:listall}
\end{minipage}

\subsection{Measured viscosity}
\label{sec:A5}
\begin{figure}
\center
\includegraphics[width=0.7\textwidth]{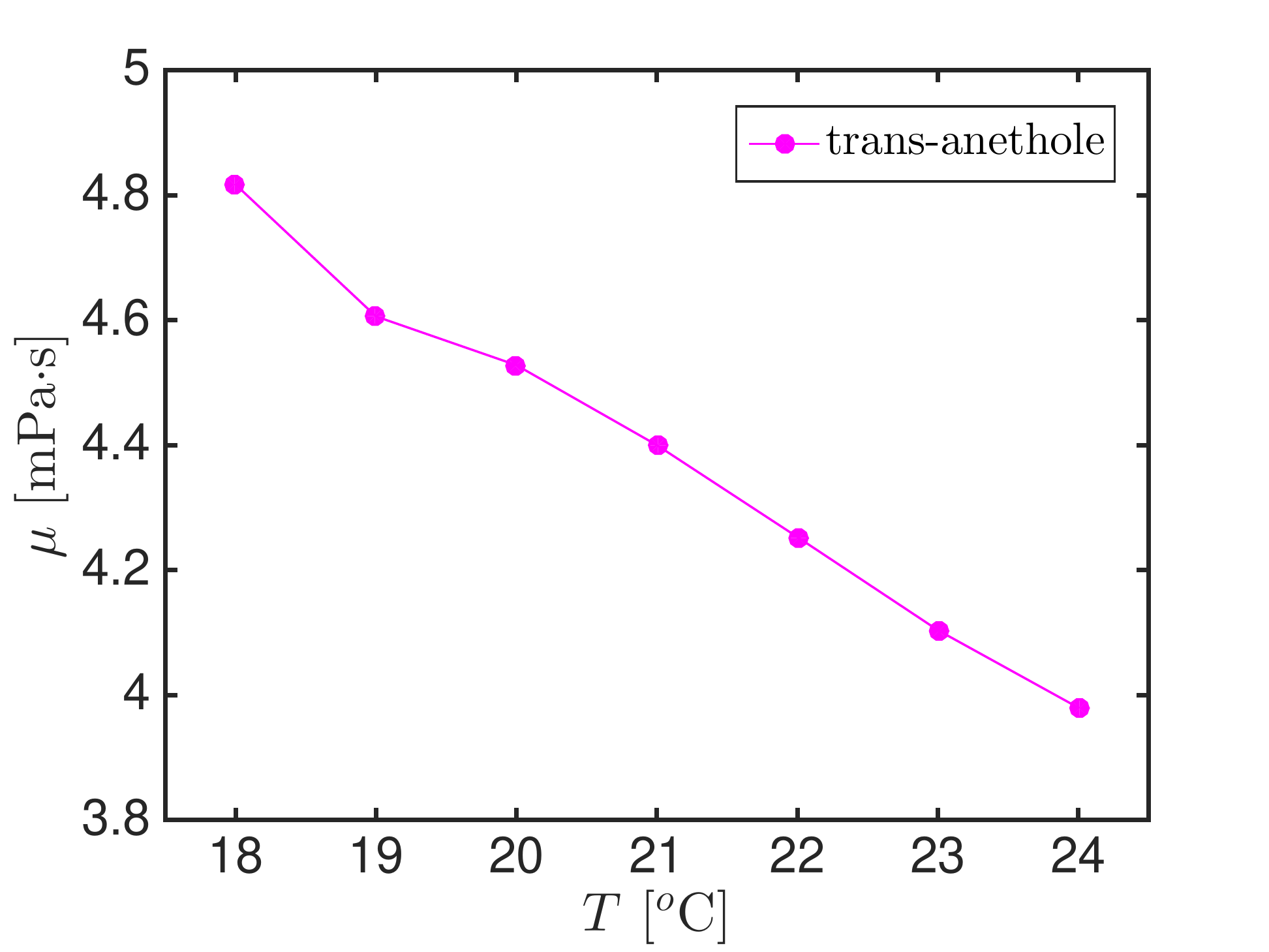}
\caption{Dynamic viscosity of anethole as function of temperature.}
\label{fig:vis}
\end{figure}

The viscosity of anethole (Fig. \ref{fig:vis}) was measured with Rheometer (Anton Paar, MCR502) at different temperatures, ranging from \SI{18}{\degree} to \SI{24}{\degree}.

\section{Symbol description}
\label{sec:B}
\newpage 
\begin{table}
  \begin{center}
\def~{\hphantom{0}}
\begin{tabular}{c|l|c}
\hline 
\textbf{Symbol} & \textbf{Description} & \textbf{unit} \\
\hline
index $\alpha$ & species: water (w); ethanol (e); anethole oil (o) & \\
superscript d & drop region & \\
superscript h & host liquid region & \\
$s$ & position of drop\&host-liquid interface & \SI{}{\meter} \\
$m_\alpha^{\text{d}}$ & mass fraction (drop region) &  \\
$m_\alpha^{\text{h}}$ & mass fraction (host liquid region) &  \\
$m_{\alpha0}^{\text{d}}$ & initial mass fraction (drop region) &  \\
$m_{\alpha0}^{\text{h}}$ & initial mass fraction (host liquid region) &  \\
$m_{\alpha s}^{\text{d}}$ & mass fraction at interface (drop region) &  \\
$m_{\alpha s}^{\text{h}}$ & mass fraction at interface (host liquid region) &  \\
$M_{\alpha s}^{\text{d}}$ & $M_{\text{a}}^{\text{d}}=\int_{-\infty}^{s} m_{\text{a}}^{\text{d}} - m_{\text{a}0}^{\text{d}} \diff x$ &  \SI{}{\meter} \\
$D_\alpha^{\text{d}}$ & diffusivity (drop region) & \SI{}{\meter^2\per\second}\\
$D_\alpha^{\text{h}}$ & diffusivity (host liquid region) &\SI{}{\meter^2\per\second} \\
$\mathcal{B}_{\alpha}^1,...,\mathcal{B}_{\alpha}^4$ & undetermined coefficients & \\ 
$\lambda$ & prefactor defined as $s/ \sqrt{t}$ & \SI{}{\meter \per \second^{\frac{1}{2}}} \\
$D_s$ & $\lambda^2$ & \SI{}{\meter^2\per\second} \\
$\mathcal{K}_\alpha$ & $\lambda / \sqrt{D_\alpha^\text{h}}$ & \\
$\mathcal{R}_\alpha$ & $\sqrt{D_\alpha^\text{h} /D_\alpha^\text{d}}$ & \\
$\chi_{\text{o}}^{\text{w}}$ & $\sqrt{D_\text{o}^\text{h} / D_\text{w}^\text{h}}$ & \\
$x_{\alpha}$ & mole fraction of species $\alpha$  & \\
$\gamma_{\alpha}$ & activity coefficient of species $\alpha$ & \\
\hline
\end{tabular}
\caption{All quantities used in the model for the ternary liquid.}
\label{tab:listall}
\end{center}
\end{table}

\section*{Acknowledgement}
We thank Ziyang Lu for providing useful information to our experiment design.
H.T. thanks for the financial support from the China Scholarship Council (CSC, file No. 201406890017). This work is part of an Industrial Partnership Programme of the Netherlands Organisation for Scientific Research (NWO), which is co-financed by Océ-Technologies B.V., University of Twente, and Eindhoven University of Technology. We also acknowledge the Netherlands Center for Multiscale Catalytic Energy Conversion (MCEC) and the \textit{Max Planck Center Twente for Complex Fluid Dynamics} for financial support.

%

\end{document}